\documentstyle[11pt,aaspp4,psfig,colordvi]{article} 

\newcommand{\kms}       {km~s$^{-1}$}

\newcommand{\etal}      {{et~al.}}
\newcommand{\apg}       {^{>}_{\sim}}

\newcommand{\vh}        {$v_{\odot}$}
\newcommand{\pcm}       {cm$^{-2}$}
\newcommand{\lya}       {Ly$\alpha$}
\newcommand{\HST}       {{\it HST\/}}
\newcommand{\WHT}       {{\it WHT\/}}

\received{19-August-1999}
\accepted{13-October-1999}

\begin{document}
\slugcomment{\em \today}

\title{\bf Interstellar and Circumstellar Optical \& Ultraviolet 
Lines Towards SN~1998S\altaffilmark{1}}

\author{David V. Bowen\altaffilmark{2,3,4}, Katherine C. Roth\altaffilmark{4},
David M. Meyer\altaffilmark{5}, \&  J. Chris Blades\altaffilmark{6}}

\author{ }

\affil{$^2$ Princeton University Observatory, Princeton, NJ 08544} 
\affil{$^3$ Royal Observatory Edinburgh, Blackford Hill, Edinburgh
EH9 3HJ, U.K.}
\affil{$^4$ Department of Physics and Astronomy, Johns Hopkins University,
Baltimore, MD 21218.}
\affil{$^5$ Department of Physics and Astronomy, Northwestern University, 
Evanston, IL 60208}
\affil{$^6$ Space Telescope Science Institute, 3700 San Martin Drive,
Baltimore, MD 21218}

\authoremail{dvb@astro.princeton.edu}

\altaffiltext{1}{Based in part on observations obtained with the NASA/ESA
Hubble Space Telescope, obtained at STScI, which is operated by the
Association of Universities for Research in Astronomy, Inc., under contract
with the National Aeronautics and Space Administration, NAS5-26555.}

\begin{abstract}

We have observed SN~1998S which exploded in NGC~3877, with the Utrecht
Echelle Spectrograph ($6-7$~\kms\ FWHM) at the {\it William Herschel
Telescope} and with the E230M echelle of the Space Telescope Imaging
Spectrograph (8~\kms\ FWHM) aboard the {\it Hubble Space Telescope}.
Both data sets were obtained at two epochs, separated by 19 (optical)
and seven days (UV data).

From our own Galaxy we detect interstellar absorption lines of
\ion{Ca}{2}~K, \ion{Fe}{2} $\lambda\lambda 2600,2586,2374,2344$,
\ion{Mg}{1} $\lambda2852$, and probably \ion{Mn}{2} $\lambda 2576$, at
$v_{\rm{LSR}}\:=\:-95$~\kms\ arising from the outer edge of the High
Velocity Cloud Complex~M. We derive gas-phase abundances of
[Fe/H]$\:=\:-1.4$ and [Mn/H]$\:=\:-1.0$, values which are very similar
to warm disk clouds found in the local ISM.  This is the first
detection of manganese from a Galactic HVC, and we believe
that the derived gas-phase abundances argue against the HVC material
having an extragalactic origin.

At the velocity of NGC~3877 we detect interstellar \ion{Mg}{1}
$\lambda 2852$, \ion{Mn}{2} $\lambda\lambda 2576, 2594,2606$,
\ion{Ca}{2}~K and \ion{Na}{1}~D$_2$,D$_1$ absorption lines, spanning a
velocity range of $-102$ to $+9$~\kms\ from the systemic velocity of
the galaxy (910~\kms ). Surprisingly, the component at $-102$~\kms\ is
seen to increase by a factor of $\apg\:1$~dex in $N$(\ion{Na}{1})
between 20-March-1998 and 8-April-1998, and in $N$(\ion{Mg}{1})
between 4-April-1998 and 11-April-1998.

Unusually, our data also show narrow Balmer, \ion{He}{1}, and
metastable UV \ion{Fe}{2} P-Cygni profiles, with a narrow absorption
component superimposed on the bottom of the profile's absorption
trough. Both the broad and narrow components of the optical lines are
seen to increase substantially in strength between March-20 and
April-8. The broad absorption covers $\sim 350$~\kms\ and is seen in
\ion{Mg}{2} $\lambda 2796,2803$ absorption as well, although there is
no evidence of narrow \ion{Mg}{2} emission forming a P-Cygni
profile. There is some suggestion that this shelf has {\it decreased}
in strength over seven days between April-4 and April-11.

Most of the low-ionization absorption can be understood in terms of
gas co-rotating with the disk of NGC~3877, providing the supernova is
at the back of the disk as we observe it, and the H~I disk is of a
similar thickness to our own Galaxy.  However, the variable component
seen in all the other lines, and the accompanying emission which
forms the classic P-Cygni profiles, most likely arise in slow moving
circumstellar outflows originating from the red supergiant progenitor of
SN~1998S.

\end{abstract}


\section{Introduction}

Supernovae which explode in galaxies other than our own Milky Way
offer a unique opportunity to study in detail an {\it extragalactic}
interstellar medium (ISM), since the supernova acts a bright
background source against which absorption from gas along the
sightline can be observed.  Supernovae have been used successfully to
probe galaxy ISMs both from the ground (e.g. \cite{wall72};
\cite{jenk84}; \cite{dodo85}; \cite{vida87}; \cite{stei90};
\cite{meye91}) using Ca~II and Na~I absorption lines, and more
recently, from space with IUE (\cite{pett82}; \cite{blad88}).  Since
the launch of \HST, we have been using the high resolution
spectrographs to study the ISMs of nearby galaxies in more
detail. Although a uniformity in the absorption line properties from
an extragalactic ISM might be expected, considering that multiple
sightlines through the Galactic ISM show very similar Mg~II
profiles, the characteristics observed have, in fact, been quite
diverse.  Towards SN~1992A, which exploded in the elliptical galaxy
NGC~1380 in the Fornax Cluster, we detected {\it no} Mg~II absorption
to sensitive equivalent width limits (\cite{bowen95}), suggesting that
the galaxy was devoid of neutral hydrogen, and that elliptical
galaxies are unlikely to contribute significantly to the numbers of
Mg~II absorption systems seen at high redshift. Conversely, toward
SN~1993J which arose in M81, we detected individual absorption
components spanning a velocity range of $\sim 400$~\kms , grouped into
3 subcomplexes arising in gas from the disk of M81, from our own
Galaxy, and from intergalactic tidal debris from interactions between
M81 and M82 (\cite{bowen94}). In this paper we add another sightline
to our sample, that of SN~1998S, and once again find unexpected
absorption line properties in the supernova's spectrum.

\begin{figure}[h]
\centerline{\psfig{figure=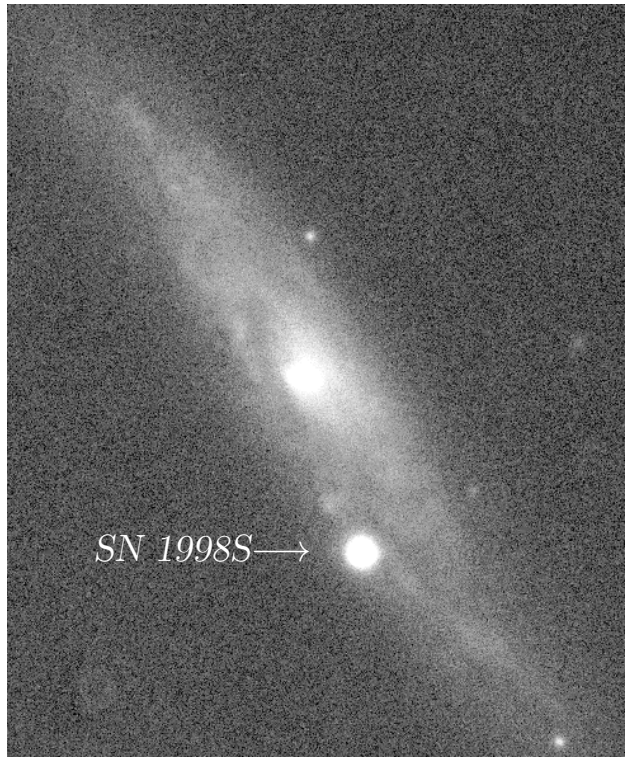,height=10cm}}
\figcaption{A 2 min R-band CCD image of NGC~3877 taken on 16-March-1998 
by S. Bennett \& A. Aragon-Salamanca at
the {\it Isaac Newton Telescope} on La Palma. North-East is
top-left. SN~1998S, lying along the spiral arm towards the
South-West, is 48 arcsec from the center of the
galaxy. \label{pic.fig}}
\vspace*{-5.2cm}\hspace*{5.3cm}\White{\Large{\sl SN 1998S$\longrightarrow$}}
\end{figure}

The type II-L supernova SN~1998S exploded in the outer regions of
NGC~3877, an edge-on Sc galaxy in the Ursa Major cluster with a
heliocentric velocity of $910\pm 6$~\kms\ (\cite{broe94}). The SN
sightline lies at a projected distance of 48 arcsec from the center of
the galaxy, or 3.6~kpc, assuming a distance of 15.5~Mpc to NGC~3877
(\cite{sand98}).  An image of the supernova and its host galaxy is
shown in Figure~\ref{pic.fig}. In this paper we present
high-resolution UV and optical spectra of the supernova, taken to
probe the ISM of the host galaxy.

\vspace*{4.5cm}

\section{Observations and Results}

\setcounter{footnote}{0}

Details of the observations are given in 
Table 1, 
which lists dates of observations, wavelength ranges covered, exposure
times, and spectral resolutions. \HST\ observations of SN~1998S were
made at two epochs, each with the Space Telescope Imaging Spectrograph
(STIS) and E230M echelle through the 0.2x0.06 slit using the NUV-MAMA
detector.  We used the orders extracted by the STSDAS pipeline
calibration; however, the subtraction of the extracted background
spectrum resulted in a poor signal-to-noise for the final spectra,
since the background count rate was a significant fraction of the
total count rate. For each order, we therefore replaced the background
spectrum with a single value obtained from averaging the data values
in the background spectra, and subtracted this from the gross
extracted counts, thereby improving the signal-to-noise of the
data. Where necessary, overlapping orders were summed after weighting
pixels by the inverse of the noise derived from the pipeline
extractions. Optical spectra were obtained with the Utrecht Echelle
Spectrograph at the {\it William Herschel Telescope}: data acquired
20-March-1998 were taken with the E31 (31.6 lines mm$^{-1}$) echelle
and a 2.0 arcsec slit, centered at 5430~\AA ; spectra taken
8-April-1998 were made with the E79 grating (79.0 lines mm$^{-1}$) and
a 1.0 arcsec slit, centered at 5900~\AA\ and 4280~\AA . For both data
sets, a SITe1 2048x2048 CCD detector was used, and both were reduced
with the NOAO {\tt echelle} IRAF package.\footnote{The Image Reduction
and Analysis Facility (IRAF) software is provided by the National
Optical Astronomy Observatories (NOAO), which is operated by the
Association of Universities for Research in Astronomy, Inc., under
contract to the National Science Foundation.  The Space Telescope
Science Data Analysis System (STSDAS) is distributed by the Space
Telescope Science Institute.}

\begin{table}[h]
\tablewidth{15cm}
\begin{center}
\begin{small}
\begin{tabular}{cllcccc}
\multicolumn{7}{c}{Table 1. Chronology of Observations}\\
\hline\hline
  & & & $\Delta\lambda$ & Exp time & FHWM & T+ \\
UT Date & Obsv. & Instrument & (\AA ) & (min) & (\kms ) 
& (days)$^a$\\
\hline
20-Mar-1998    & \WHT & UES    & 4190$-$6610  & \phn20  & 6.0 & \phs\phn0  \\
4-Apr-1998     & \HST & STIS+E230M      & 2277$-$3115  & 195     & 8.3 & +15 \\
8-Apr-1998     & \WHT & UES    & 3620$-$5860  & \phn60  & 7.4 & +19 \\
 	       &     &                 & 4530$-$9050  & \phn60  & 7.4 &    \\
11-Apr-1998    & \HST & STIS+E230M      & 2277$-$3115  & 349     & 8.3 & +22 \\
\hline
\multicolumn{7}{l}{$^a$ Days after maximum light, assumed to be 20-March-1998. Discovery was 3-March-1998.}
\end{tabular}
\end{small}
\end{center}
\end{table}

\subsection{ Cold interstellar gas in the Milky Way \& from HVC Complex M}

Fortuitously, at Galactic co-ordinates of $l\:=\:150.7$, $b\:=\:66.0$,
the sightline to SN~1998S intercepts the group of High Velocity Clouds
(HVCs) designated as MI in Complex~M (\cite{wakk91}). The complex is
thought to lie at a height of $\leq\:3.5$~kpc above the Galactic plane
(\cite{ryan97}; \cite{danl93}; \cite{keen95}) and is the highest
latitude conglomerate of HVCs known. In fact, the sightline passes
through the very edge of the complex (as measured at 21~cm), but data
from the Leiden/Dwingeloo 21~cm Survey (\cite{hart97}) show that there
exists strong emission at $v_{\rm{LSR}}\:\simeq\:-87$~\kms , at least
over the 36 arcmin beam centered at $l\:=\:150.5$, $b\:=\:66.0$.

\begin{figure}
\psfig{figure=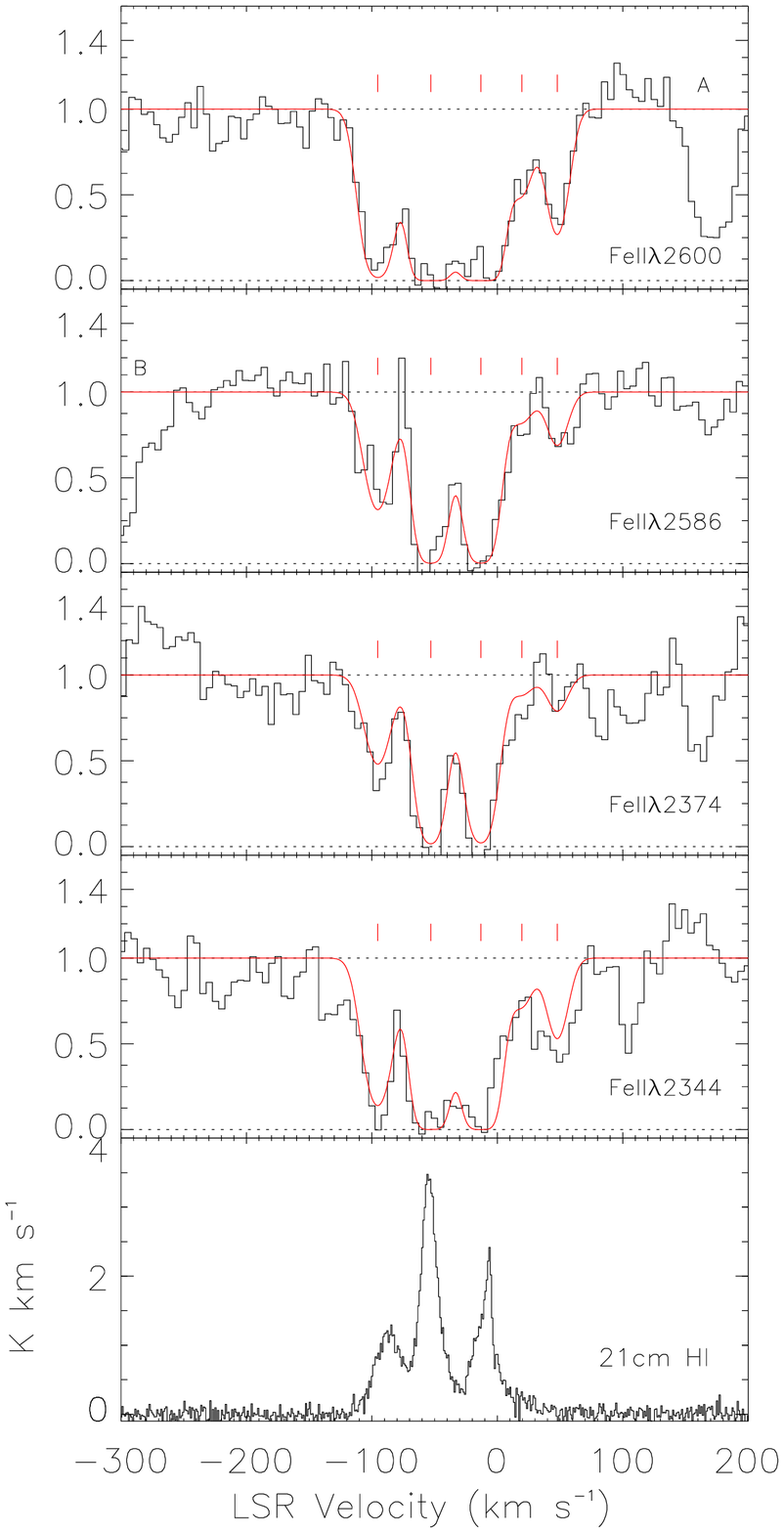,height=16cm,angle=0}
\vspace*{-16cm}\hspace*{8cm}\psfig{figure=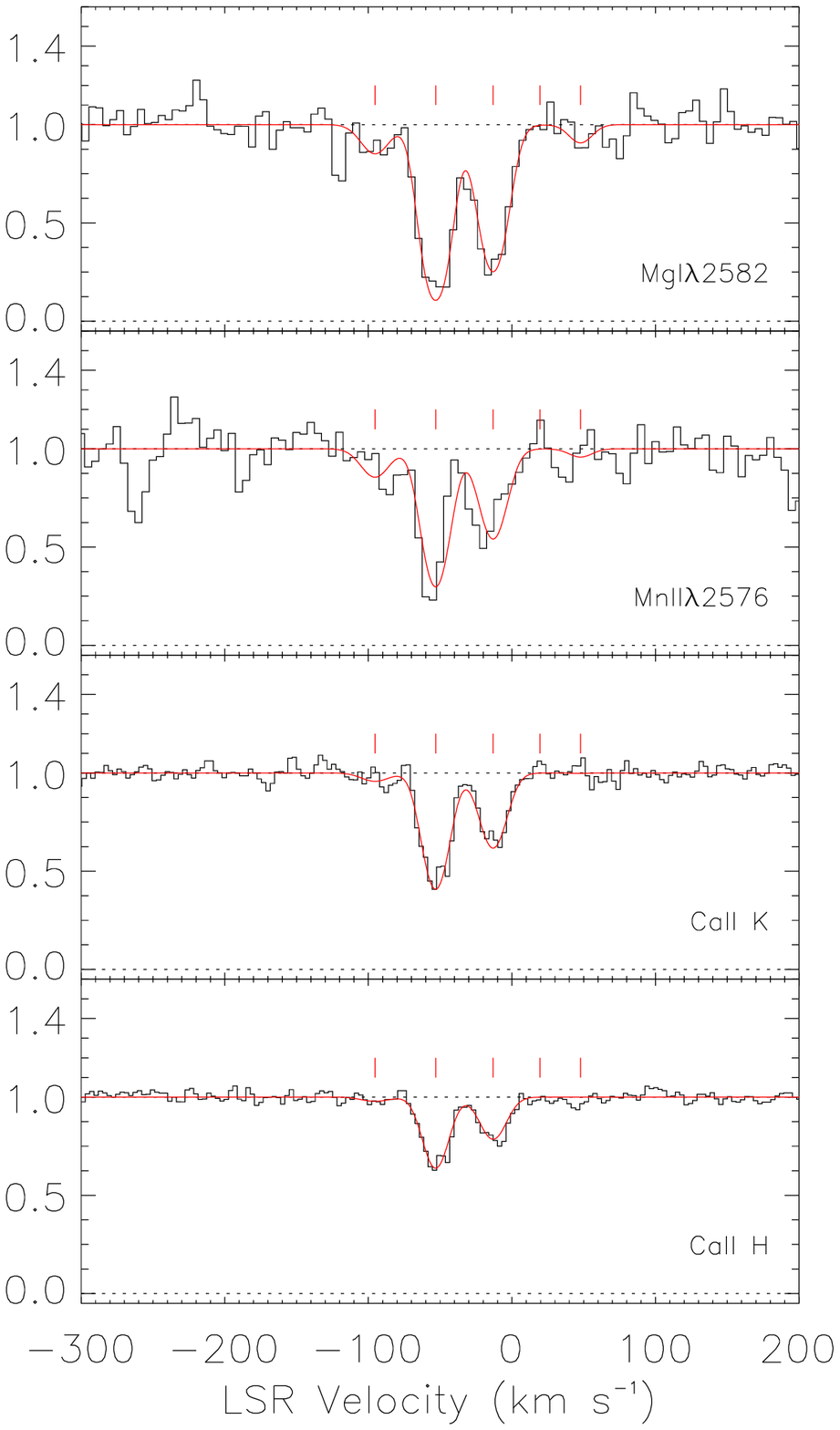,height=16cm,angle=0}
\figcaption{Interstellar Fe~II (a, left), Mg~I, Mn~II (b, right), and
Ca~II~K, H absorption from the Milky Way along the sightline to
SN~1998S.  The values of $v_{\rm{LSR}}$ derived from Voigt profile
fits are indicated by tick marks above the best fits themselves (solid
line). The line labelled 'A' is Mn~II $\lambda 2594$ arising in
NGC~3877, and that labelled 'B' is Mn~II $\lambda 2576$ also in the
host galaxy.
Absorption can be seen at a velocity of
$v_\odot\:=\:-101$~\kms\ ($v_{\rm{LSR}}\:=\:-95$~\kms ), which arises
from the edge of the High Velocity Cloud Complex~M.  The 21~cm
emission along the sightline, centered at $l\:=\:150.5$, $b\:=\:66$,
from the Leiden/Dwingeloo H~I Survey (Hartmann \& Burton 1997) is
shown bottom of (a). \label{hvc_1.fig}}
\end{figure}

In our STIS spectra, we detect \ion{Fe}{2} $\lambda\lambda 2600, 2586,
2374, 2344$, \ion{Mg}{1} $\lambda 2852$, and, probably,
\ion{Mn}{2} $\lambda2576$ absorption lines at
$v_{\rm{LSR}}\:=\:-95$~\kms\ (Figure~\ref{hvc_1.fig}). This velocity
agrees well with the 21~cm emission from Complex~M, which we reproduce
at the bottom of Figure~\ref{hvc_1.fig}a. This is the first
detection of manganese in an HVC (\cite{wakk97}).  The
\ion{Fe}{2} $\lambda 2382$ line from the Milky Way is also covered by
our STIS spectra, but the HVC absorption is blended with redshifted
\ion{Fe}{2} $\lambda 2374$ from NGC~3877 and is disregarded in our
analysis.  Similarly, \ion{Mn}{2} $\lambda2594$ and
\ion{Mn}{2} $\lambda2606$ absorption which might arise from the HVC are
blended with other redshifted \ion{Fe}{2}
lines. Figure~\ref{hvc_1.fig}b shows that in the optical, weak
\ion{Ca}{2}~K is also detected (at a statistically more significant
level that the Mg~I, since the signal-to-noise of the optical data is so much
higher than the STIS data), but the \ion{Ca}{2}~H line is absent.
Galactic \ion{Na}{1}~D$_1$,D$_2$ is badly contaminated by
\ion{He}{1} $\lambda5877$ absorption and emission at the redshift of
NGC~3877 (see \S2.3), making it impossible to derive reliable
measurements of absorption from the HVC. Equivalent widths, $W_\lambda$, and
errors, $\sigma$($W_\lambda$), to the HVC
lines are given in Table~2, derived from standard methods
(e.g. Bowen~\etal~1995 and refs. therein)
--- the \ion{Mn}{2} $\lambda2576$
is detected at exactly the 3$\sigma$ level. The \ion{Fe}{2} lines from
the HVC are heavily blended with lower velocity absorption, so their
measured equivalent widths are given as lower limits in
Table~2.


As can be seen in Figure~\ref{hvc_1.fig}, we also detect \ion{Fe}{2} and
possibly \ion{Mg}{1} at $v_{\rm{LSR}}\:=\:+48$~\kms . These lines are
highly unusual given that there is no 21~cm emission seen at this
velocity. The absorption must arise either in gas with unusual
ionization conditions or else must represent dense gas whose size is
very much smaller than the width of the 21~cm beam, and whose
contribution to the emission is therefore too weak to be detected.

To estimate column densities of these clouds, we used our absorption line
fitting routine {\tt MADRIGAL} (\cite{bowen95}) to best fit
theoretical Voigt profiles to \ion{Fe}{2} $\lambda\lambda 2600, 2586,
2374, 2344$, \ion{Mn}{2} $\lambda 2576$, \ion{Mg}{1} $\lambda 2852$, and
\ion{Ca}{2} H,~K lines simultaneously. We used the sum of the April-4
and April-11 STIS data to improve the signal-to-noise of the UV lines,
and adopted a Gaussian Line Spread Function (LSF) with a Doppler width of
$b_{\rm{instr}}\:=\:5.0$~\kms\ for the E230M echelle, as described in
the STIS Instrument Handbook. We assumed that all ions would have a
single value of $v_{\rm{LSR}}$ and Doppler width $b$; the values of
$b$ for individual ions would be different if the widths of the lines
are dominated by kinetic temperatures rather than bulk turbulent
motions (since $b\:=\:[(2kT/m) + 2v_t^2]^{1/2}$, where $m$ is the
atomic mass of the species, $T$ the kinetic temperature of the gas,
and $v_t$ represents the turbulence in the gas). However, the lines
are clearly resolved in our data, show no obvious substructure, and
have $b$ values which would not be expected for the low-ionization
lines seen if representative of kinetic temperatures. It therefore
seems likely that the width of the lines are dominated by turbulent
motions, and we adopted a single $b$ value for all species.

The data are of insufficient quality to properly derive column
densities for the absorption from the Milky Way near zero velocity;
nevertheless, since absorption at these velocities contributes to the
higher velocity lines, we need to include lower velocity components,
even though the fits will not be precise.  The results are given in
Table~2, and fits to the \ion{Fe}{2}, \ion{Mg}{1} and
\ion{Ca}{2} lines are shown in Figure \ref{hvc_1.fig}.  The values of
$v_{\rm{LSR}}$ are indicated by tick marks.  We find a velocity of
$v_{\rm{LSR}}\:=\:-95$~\kms\ for the HVC, and derive $b\:=\:12.4$~\kms
, which agrees well with a simple Gaussian fit to the 21~cm emission
feature shown in Figure \ref{hvc_1.fig}a,
$b_{\rm{21cm}}\:=\:15.0$~\kms .  For the cloud at positive velocities
we measure $v_{\rm{LSR}}\:=\:+48$~\kms , $b\:=\:10.0$~\kms .  We have
derived errors in the column densities by re-fitting 500 synthetic
lines generated from the initial fit to the data, each spectrum given
the same signal-to-noise as the original data (see
Bowen~et~al.~1995). The errors quoted in Table~2 refer to
the value of $\sigma$ derived from a Gaussian fit to the distribution
of $b$ and $N$ measured from the synthetic data, and hence refer only
to errors associated with the noise in the spectra. Errors in both $b$
and $v_{\rm{LSR}}$ arising from these Poisson statistics are small,
$0.3$~\kms . In reality, the error in $v_{\rm{LSR}}$ will be dominated
by the wavelength calibration of the STIS spectrum, while the value of
$b$ will be affected by uncertainties in the STIS LSF. In the first
case, the error in the dispersion solution and zero-point for the
MAMAs is no more that $\sim 1$ pixel, or $\sim 5$~\kms , while for the
optical data, the residuals of the fits to the echelle arc lines
suggests an error of only $\sim 0.2$~\kms . In the second case, the
accuracies of the STIS LSFs still remain largely unexplored.

\begin{table}[h]
\begin{center}
\begin{small}
\tablewidth{18cm}
\begin{tabular}{lcrrrccrrlc}
\multicolumn{11}{c}{Table 2. Column densities towards Galactic Clouds}\\
\hline\hline
 &&\multicolumn{4}{c}{Complex M} & &\multicolumn{4}{c}{}\\
 &&\multicolumn{4}{c}{$v_{\rm{LSR}}\:=\:-95.3$ \kms,
 $b\:=\:12.4$ \kms} & & \multicolumn{4}{c}{$v_{\rm{LSR}}\:=\:+47.8$
\kms, $b\:=\:10.0$ \kms}\\
\cline{2-6}\cline{8-11}
              && $W_\lambda$ $^a$    & $\sigma$($W_\lambda$) &        & 
&& $W_\lambda$ $^a$    & $\sigma$($W_\lambda$) &        &  \\
\phm{000} Ion && (m\AA )                 & (m\AA )       &    $\log N$  & $\sigma$($\log N$)
&& (m\AA )                & (m\AA )       &    \phm{..}$\log N$  & $\sigma$($\log N$) \\
\hline
\ion{Mn}{2} $\lambda2576$    && $\leq 39$  & 13
& $\leq\:12.28$             & 0.11 &  & $< 24 $ & \nodata & $<12.2$ & ... \\

\ion{Fe}{2} $\lambda2599$    && 300     & 15
&  13.85                    & 0.01 &   & 110      & 6 & \phm{1}\phm{..}13.27 & 0.05 \\

\ion{Fe}{2} $\lambda2586$    && 138     & 8
&                           &      &   & 60       & 9 &  &  \\

\ion{Fe}{2} $\lambda2374$    && 140     & 25
&                           &      &   & 97       & 15 & &  \\

\ion{Fe}{2} $\lambda2344$    && 204	   & 10
&                           &      &   & 159        & 12 & &  \\

\ion{Mg}{1} $\lambda2582$    && 29         & 8
&  11.45                    & 0.09 &  & $\leq 13$ & 6 & $\leq11.2$ & ... \\

\ion{Ca}{2}~K               && 13         & 3
&  11.15                    & 0.12  &  & $< 6 $ & ... & $<11.1$ & ...\\

\ion{Ca}{2}~H               && $< 6$     & \nodata
&                           &        &  & $< 6$ & ... &  & \\ 
\hline
\multicolumn{11}{l}{$^a$Measured values are lower limits since the lines are
blended with components nearer zero \kms .}\\
\multicolumn{11}{l}{Limits are 3$\sigma$($W_\lambda$)}
\end{tabular}
\end{small}
\end{center}
\end{table}

\begin{figure}[h]
\vspace*{-1.5cm}\centerline{\psfig{figure=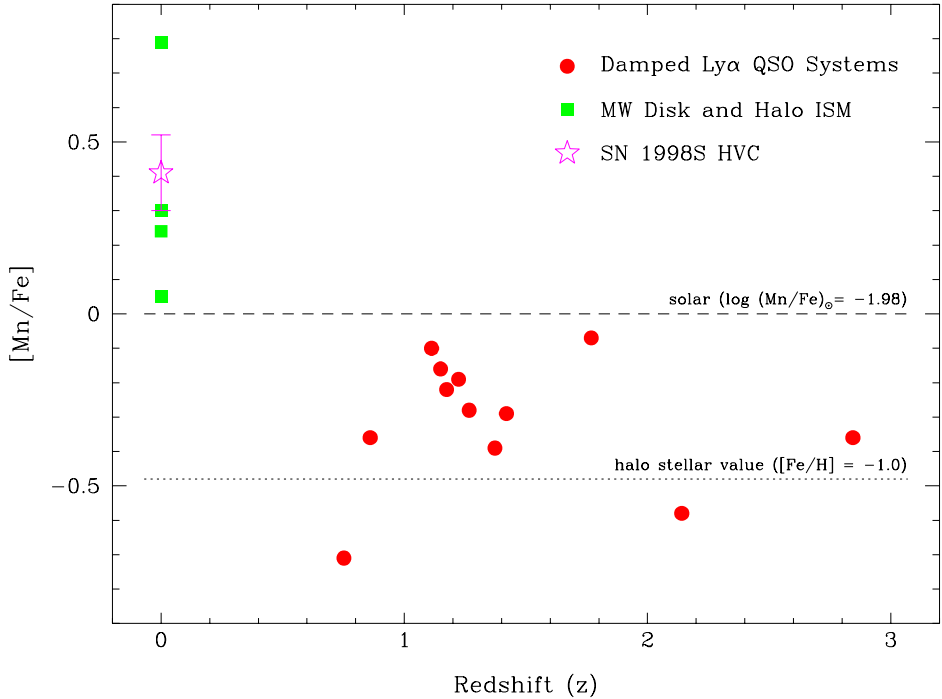,height=12cm,angle=270}}
\vspace*{-1cm}\figcaption{[Mn/Fe] values in the Galactic ISM and damped Ly$\alpha$
absorption systems plotted as a function of redshift.  Our marginal
Mn~II detection toward the HVC in the SN 1998S sightline infers
the [Mn/Fe] value indicated by a star symbol.  The solar abundance
ratio (Mn/Fe)$_\odot$ marked by a dashed line is from the meteoritic
abundances of Anders \& Grevesse (1989).  Galactic halo star measurements are
also shown, taken from \cite{rya96}.  Measurements of [Mn/Fe] for
damped Ly$\alpha$ QSO absorbers can be found in \cite{lu96},
\cite{lu95}, \cite{mey95}, \cite{pet99}, \cite{rot99}, and
\cite{ste95}. \label{MnFe.fig}}
\end{figure}

With no detection of H~I towards the cloud at
$v_{\rm{LSR}}\:=\:+48$~\kms , we can derive little about the physical
conditions of the absorbing gas. However, for the absorption by the
HVC in Complex~M, we can make crude estimates of the metallicity of
the HVC.  The \ion{H}{1} column density estimated from the
Leiden/Dwingeloo 21~cm Survey is $\simeq\:6\times10^{19}$~\pcm
. Adopting the cosmic meteoritic abundance ratios $\log ({\rm Fe}/{\rm
H})_\odot=-4.49$ and $\log ({\rm Mn}/{\rm H})_\odot=-6.47$
(\cite{and89}), we derive gas-phase depletion values of Fe and Mn
$-1.44$ and $-1.03$ respectively.\footnote{$[{\rm X}/{\rm H}]\equiv
\log ({\rm X}/{\rm H}) - \log({\rm X}/{\rm H})_\odot$} In the Galactic
interstellar medium both Mn and Fe are heavily depleted onto dust
grains, and in fact, the values of [Fe/H] and [Mn/H] are remarkably
similar to those seen in the local `warm disk' clouds discussed by
\cite{sav96}.  However, dust depletion is not the only explanation for
sub-solar gas-phase depletions --- a combination of low metallicity
and less dust depletion within the HVC cloud could give similar
values.  Stellar abundance measurements toward moderately metal-poor
($[$Fe$/$H$]=-1.0$) halo stars find Mn underabundant with respect to
Fe with an average $[$Mn$/$Fe$]=-0.48$ (Ryan, Norris, \& Beers 1996).
High redshift measurements in damped Ly$\alpha$ QSO absorption systems
find $[{\rm Mn}/{\rm Fe}]$ values consistently below zero, which has
been interpreted as resulting from a combination of a warm Galactic
disk-like dust depletion pattern superposed on a slightly sub-solar
Mn$/$Fe abundance ratio (\cite{vla98}).  We present the ISM, stellar,
damped Ly$\alpha$ QSO $[{\rm Mn}/{\rm Fe}]$ measurements, and our new
HVC value, in Figure~\ref{MnFe.fig}. The derived [Fe/H] and [Mn/H]
ratios give a gas-phase depletion value of $[{\rm Mn}/{\rm
Fe}]=0.41\pm0.11$ for the HVC, quite different from the values in
metal-poor stars or extragalactic D\lya\ systems.


Although not conclusive, (since the point may only be an upper limit
if the detection of \ion{Mn}{2} is erroneous) our data are consistent
with the HVC material at $v_{\rm LSR}=-100$ km s$^{-1}$ having a solar
Mn$/$Fe abundance ratio and a dust depletion pattern similar to that
observed in warm disk clouds in the Galactic ISM.  This interpretation
argues against the HVC material having an extragalactic origin as has
been proposed for some HVCs (e.g. \cite{bli99}).

\subsection{Cold interstellar gas in NGC3877}

\begin{figure}
\psfig{figure=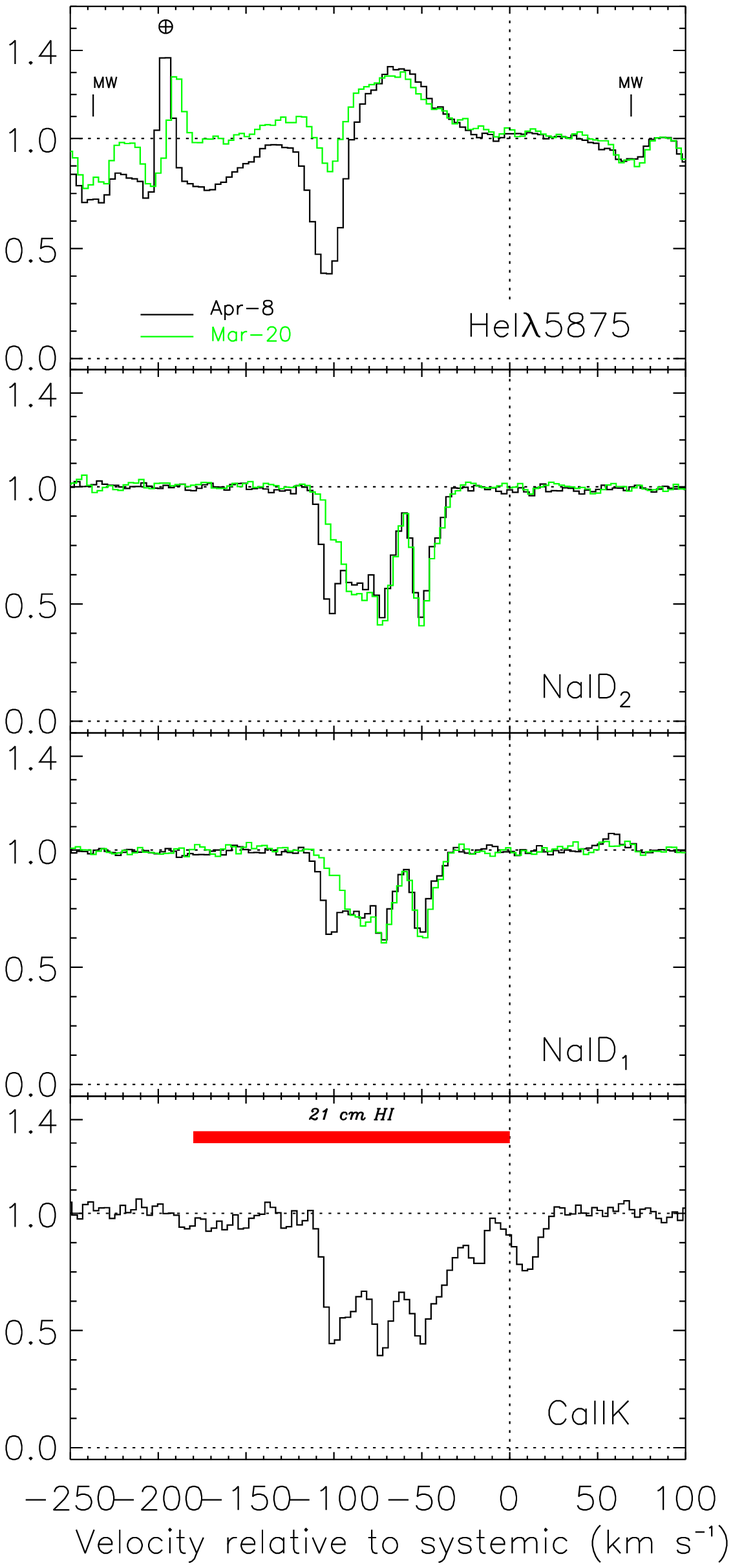,height=17cm,angle=0}
\vspace*{-17cm}\hspace*{8cm}\psfig{figure=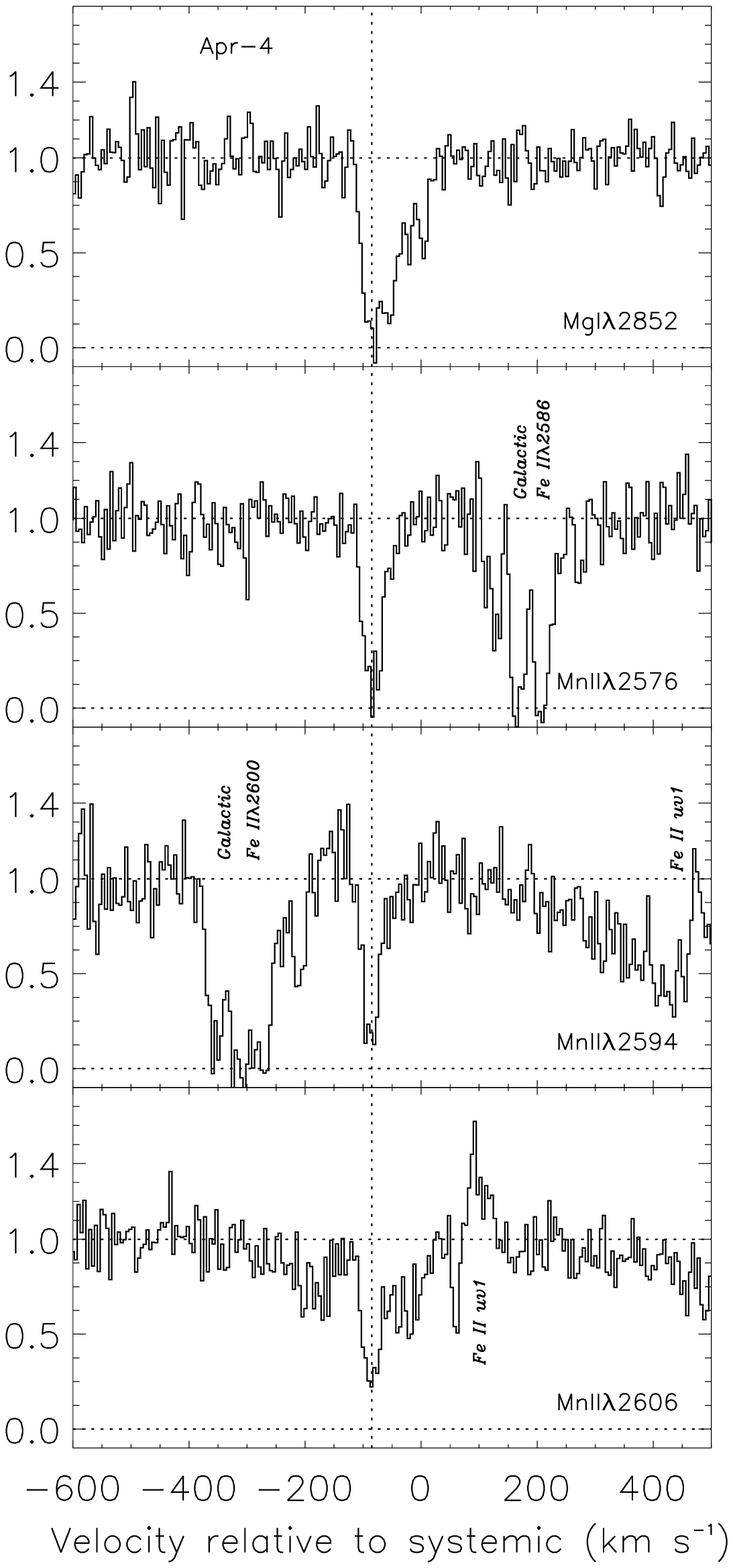,height=17cm,angle=0}
\figcaption{a) (left) Plot of He~I $\lambda5875$, and interstellar
Na~I and Ca~II~K
lines vs. $v-v_{\rm{systemic}}$ arising in NGC~3877 obtained with the
UES at the \WHT, taken between 20-Mar-1998 (heavy line) and 8-Apr-1998
(light line).  The
Na~I $\lambda5889$ night sky line is marked by a $\oplus$. The position
of Na~I absorption from the Milky Way is also indicated in the top
panel. A narrow Na~I component at $-102$~\kms\ varies
over 19 days; this is also seen in He~I absorption, which
varies over the same timescale. The development of a broad wing
in He~I absorption at $v-v_{\rm{systemic}}\:<\:-100$~\kms\ is also
apparent. Only one set of data covering Ca~II were available, so
variability in that species is undetected. The total extent of H~I
observed
at 21~cm is represented by a bar in the bottom panel, while the dotted
vertical line represents the systemic velocity of the host galaxy.
b) (right) Plot of interstellar Mg~I and Mn~II arising in NGC~3877,
taken from April-4's STIS data. 
\label{vel1b.fig}}
\end{figure}

The optical, and UV data in particular, show a wealth of absorption
lines arising from NGC~3877, and in this section we focus on the
low-ionization species detected. Figure~\ref{vel1b.fig}a shows the
optical \ion{Na}{1} and \ion{Ca}{2}~K lines, with the abscissa showing
the velocity of the components relative to the systemic velocity of
the galaxy (taken to be 910~\kms ). \ion{Ca}{2}~H is clearly detected
at the velocity of NGC~3877, but is contaminated by H$\epsilon$
emission \& absorption (see \S2.3). The absorption spans a range of
velocities covering a total spread of $\sim\:60$~\kms\ for \ion{Na}{1}
and $\sim\:100$~\kms\ for \ion{Ca}{2}, each complex clearly made up of
a number of components. The range of absorption is consistent with the
observed \ion{H}{1} emission along the sightline, as deduced from the
21~cm emission maps of Broeils \& van Woerden (1994), and marked by a
bar in the bottom panel.  We have plotted data taken at the two
different epochs (\ion{Ca}{2} was only observed April-8); the most
striking thing about the difference between the two profiles is the
increase in the most negative component of the \ion{Na}{1} absorption
over an interval of 19 days. This increase in absorption corresponds
to an increase in a narrow absorption line at the bottom of the
absorption trough of the \ion{He}{1} $\lambda5875$ P-Cygni
profile. Although these profiles are discussed in more detail in
\S2.3, we introduce this feature, shown in the top panel of
Figure~\ref{vel1b.fig}a, to highlight the unusual nature of the
variable \ion{Na}{1} component.


In Figure~\ref{vel1b.fig}b we show absorption line profiles of
\ion{Mg}{1} $\lambda 2852$ and the \ion{Mn}{2} triplet centered on
\ion{Mn}{2} $\lambda 2594$. We show only April-4's data for the sake
of clarity, but there is evidence that the strength of the blue wing
of the \ion{Mg}{1} absorption has increased, which would correspond to
the same increase seen in the \ion{Na}{1} component. The same may true
for the \ion{Mn}{2} lines, but the increase in each of the multiplet
lines is small, and within the variations from the low signal-to-noise
of the data. We note too that \ion{Fe}{2} $\lambda\lambda2600,2586$
and \ion{Fe}{2} $\lambda\lambda2344,2374,2382$ are also detected at
the velocity of NGC~3877 in the STIS data. However, the lines are
heavily blended with either zero-redshift Milky Way absorption from
various species, or metastable \ion{Fe}{2} absorption from the host
galaxy. For these reasons we choose not to include the lines in our
analysis.

\subsubsection{Derivation of column densities and Doppler parameters}

To derive column densities, $N$, and Doppler parameters, $b$, of the
low-ionization gas, we applied {\tt MADRIGAL} to the
\ion{Na}{1}~D$_2$,D$_1$ and \ion{Ca}{2}~K lines taken 8-Apr-1998.  A
minimum of nine components were required. We initially attempted to
fit both optical and UV absorption lines simultaneously. However, it
was clear that the lower resolution and lower signal-to-noise of the
STIS data meant that the spectra offered few data points with which to
constrain the fits. Most of the information on the velocity structure
in the extragalactic absorption is only evident in the optical data.
We therefore first used a simultaneous fit to the \ion{Ca}{2} and
\ion{Na}{1} complexes to derive velocities for individual components.
Since the atomic masses of both calcium and sodium are different, in
principle, the values of $b$ derived from a fit need not be the same
for each ion if the line widths are dominated by the kinetic
temperature of the gas. Hence with \vh\ constrained, we refit both the
\ion{Ca}{2} and \ion{Na}{1} allowing $b$ \& $N$ to vary for most of
the components. (Obviously, we assume that \vh\ is the same for each
ion, although this assumption depends on the geometry and physical
conditions in the cloud.) 
In fact, the \ion{Ca}{2} components are
poorly determined without the availability of the \ion{Ca}{2}~H line;
a fit to both \ion{Ca}{2} and \ion{Na}{1} simultaneously, with $b$ set
to be the same for both species (but varying during the fit), merely
results in nearly the same $b$ values derived for \ion{Na}{1}, along
with adjusted column densities for \ion{Ca}{2}. We thus re-considered
the values of $N$(\ion{Ca}{2}) derived by using only the $b$
values from \ion{Na}{1}, but found little change in the
resulting \ion{Ca}{2} column densities.

\begin{figure}
\psfig{figure=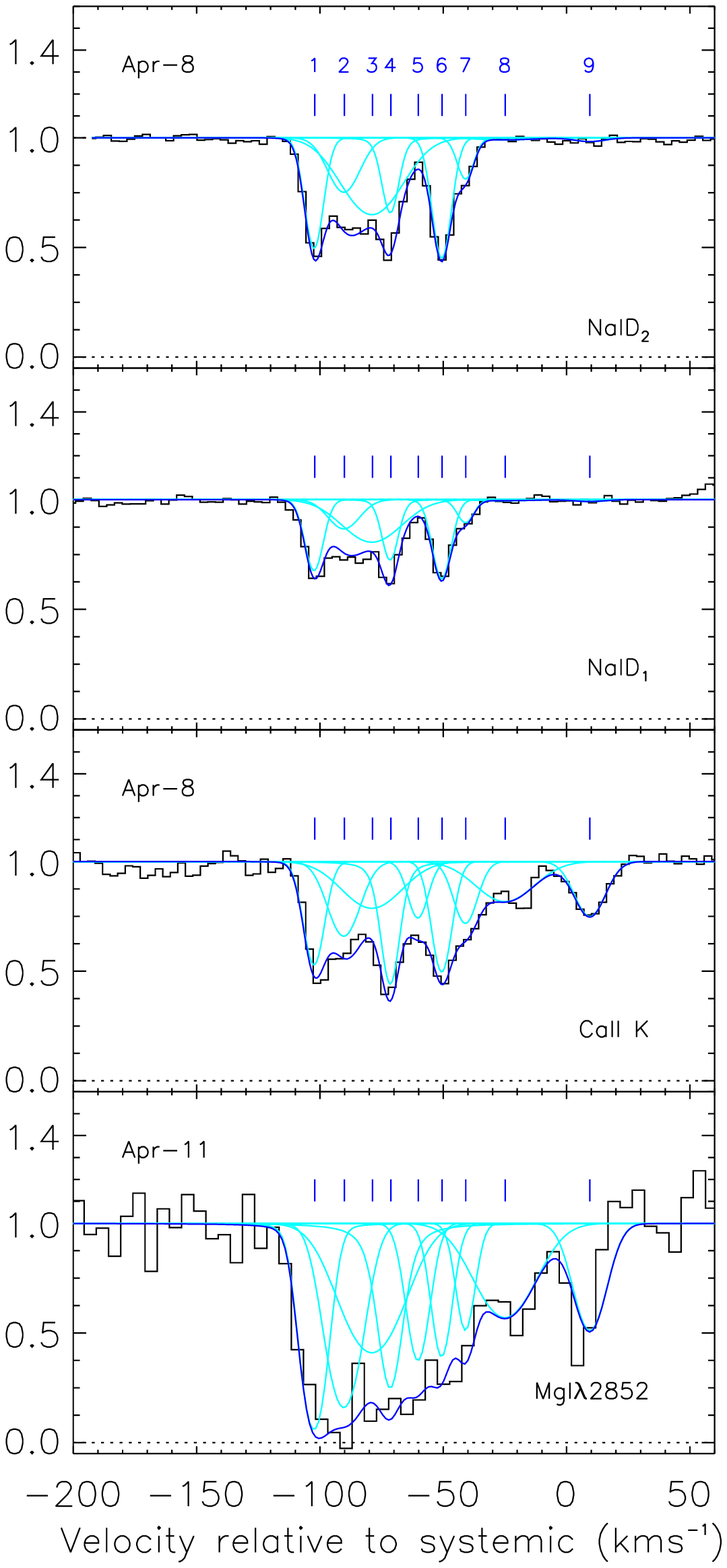,height=18cm,angle=0}
\vspace*{-18cm}\hspace*{8cm}\psfig{figure=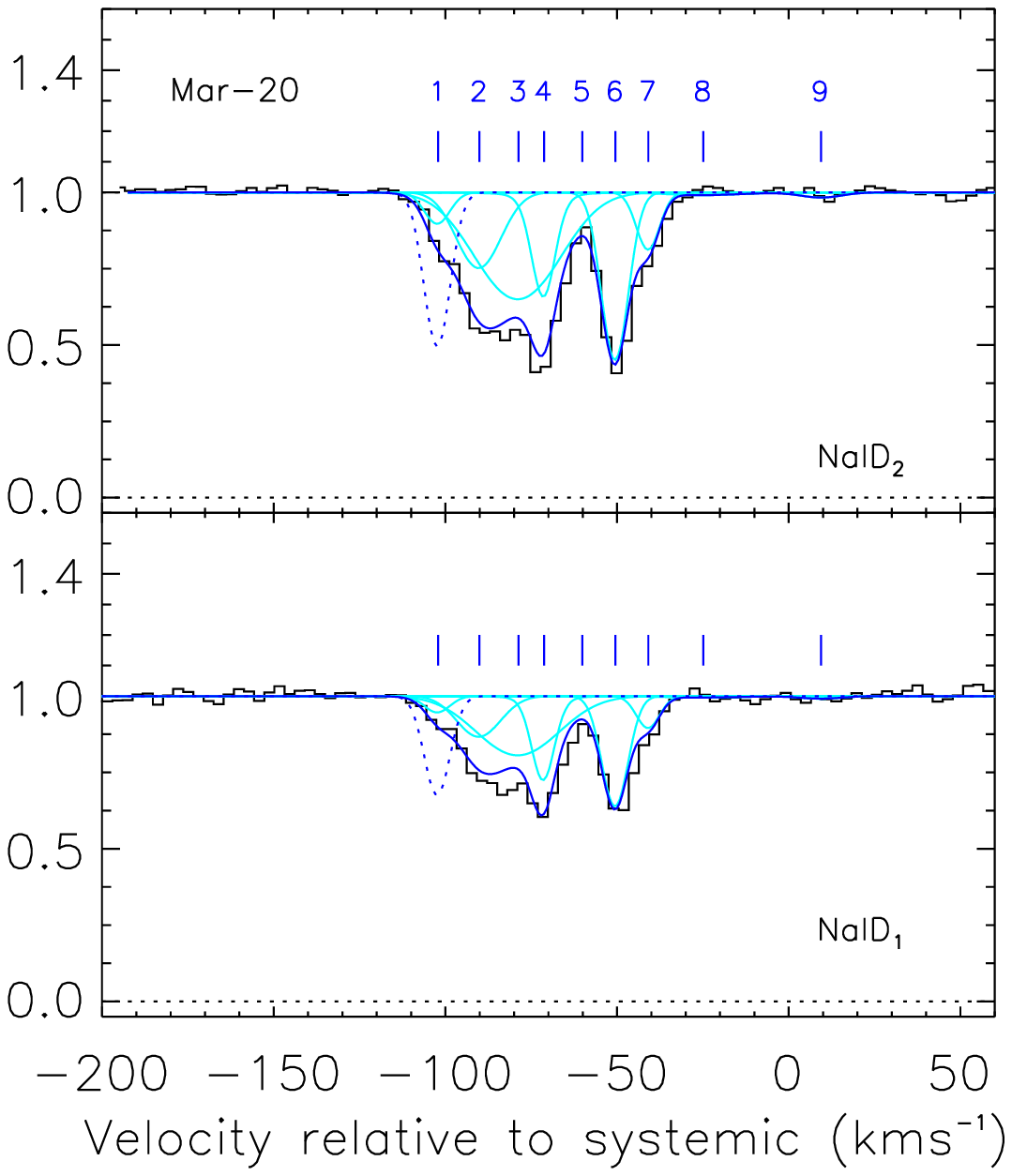,height=18cm,angle=0}
\figcaption{
{\it a}) (left) Fits of theoretical Voigt profiles to 8-April's optical and 11-April's
UV data. The solid line
shows the resulting line profile from a convolution of 9
components. The lighter lines show the profile of each component for
the values given in Table~3, to highlight how each contributes to 
the final convolved profile. 
b) (right) Fits of theoretical Voigt profiles to 20-March's optical
data. The values of $b_{\rm{Na~I}}$ and $N$(Na~I) are the same as
8-April's data shown in (a) except for component 1, which has
clearly increased in column between the two epochs. The dotted line
represents 8-April's fit to component 1 to highlight the change.
\label{exg.fig}}
\end{figure}

With plausible values of $b$ and $v_{\odot}$ established for
\ion{Na}{1} and \ion{Ca}{2}, we then fit the \ion{Mg}{1} $\lambda
2852$ and \ion{Mn}{2} $\lambda\lambda 2576,2594$ lines taken with STIS
on Apr-11. Magnesium has an atomic mass next to sodium, so we kept
\vh\ and $b$ fixed to the values derived from the \ion{Na}{1} fits ---
again, only the {\it depth} of the \ion{Mg}{1} line is constrained in
the line profile, and fixed $b$ values are necessary to produce a
meaningful fit. Manganese has an atomic mass closer to calcium, so we
fit the \ion{Mn}{2} line using the $b$ values derived for the
\ion{Ca}{2} lines. The fits to the manganese lines have the advantage
that we used two lines of the multiplet, which help constrain $b$ and
$N$, but the low resolution and poor signal-to-noise still make the
fits uncertain.  The fits are shown in Figure~\ref{exg.fig}a.


$N$ and $b$ were then derived for \ion{Na}{1}~D$_2$,D$_1$ data taken
March-20.  We found that we were able to well fit the data using
exactly the same values of \vh , $b$, and $N$(\ion{Na}{1}) derived
from April's data, except for component 1, where $N$(\ion{Na}{1})
appears to have varied by nearly 1 dex in the space of 19 days. The
profiles are shown in Figure~\ref{exg.fig}b. Finally, we refit
4-April's \ion{Mg}{1} again using the values derived from April's data
but allowing $N$(\ion{Mg}{1}) to vary for component~1.  Formally, our
fits suggest that $N$(\ion{Mg}{1}) has increased by a factor of $\sim
200$.  In these data, however, the component is close to being
saturated, and with the low signal-to-noise of the data, large
increases in $N$ can occur with very small variations in the noise of
the line. Nevertheless, some constraint on the fit comes from the fact
that the line has widened slightly as well as increased in
depth. Hence although $N$(\ion{Mg}{1}) may be over-estimated, an
increase over $1-2$ dex still seems probable. All the derived physical
parameters are given in Table~3. 

\begin{table}
\vspace*{-2cm}
\hspace*{-6cm}
\psfig{figure=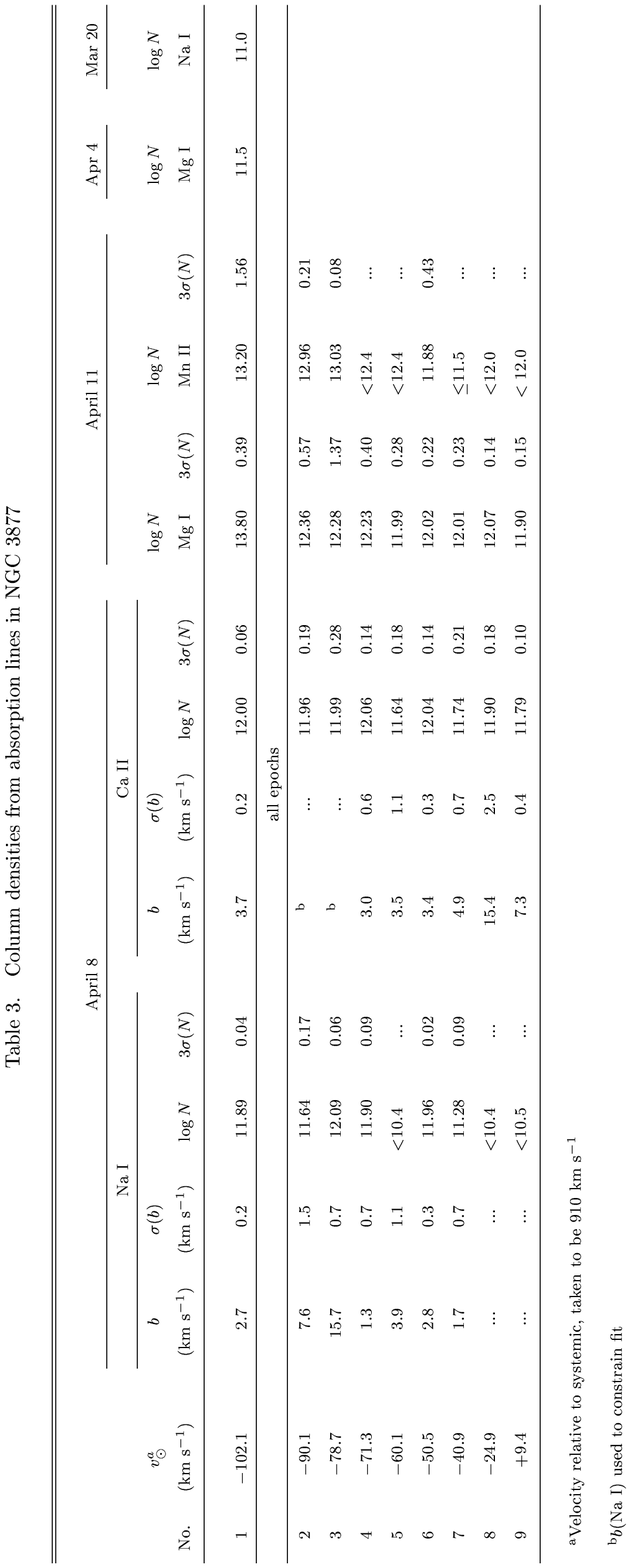,angle=180,height=25cm}
\end{table}

Ignoring the variable component (which contributes $\sim 0.1$~\AA ),
the equivalent widths of the optical lines seen in NGC~3877 are
$W_\lambda$(Ca~II$\lambda 3933$)$\:=\:0.524\pm0.007$, $W_\lambda$(Na~I$\lambda
5889$)$\:=\:0.499\pm0.005$, and $W_\lambda$(Na~I$\lambda
5895$)$\:=\:0.319\pm0.005$ (\ion{Ca}{2}~H is blended with H$\epsilon$
from the host galaxy). The \ion{Na}{1} equivalent widths are entirely
consistent with the distribution measured along sightlines to O- and
B-type stars in the Galaxy (\cite{semb93}) while the \ion{Ca}{2}
equivalent widths are slightly higher than the median of the
equivalent width distribution seen along extragalactic sightlines
through the Milky Way [0.1$-$0.2~\AA\ for \ion{Ca}{2}
(\cite{bowen91})]. Similarly, the total column densities measured in
NGC~3877 [$\Sigma \log N$(Na~I)$\:=\:12.6$, $\Sigma \log
N$(Ca~II)$\:=\:12.8$], again excluding the variable component, are
close to those measured through the entire length of the Galactic
plane (\cite{semb94}). In this sense, the optical lines suggest that
the ISM of NGC~3877 along the line of sight is
similar to our own.

\begin{figure}
\centerline{\psfig{figure=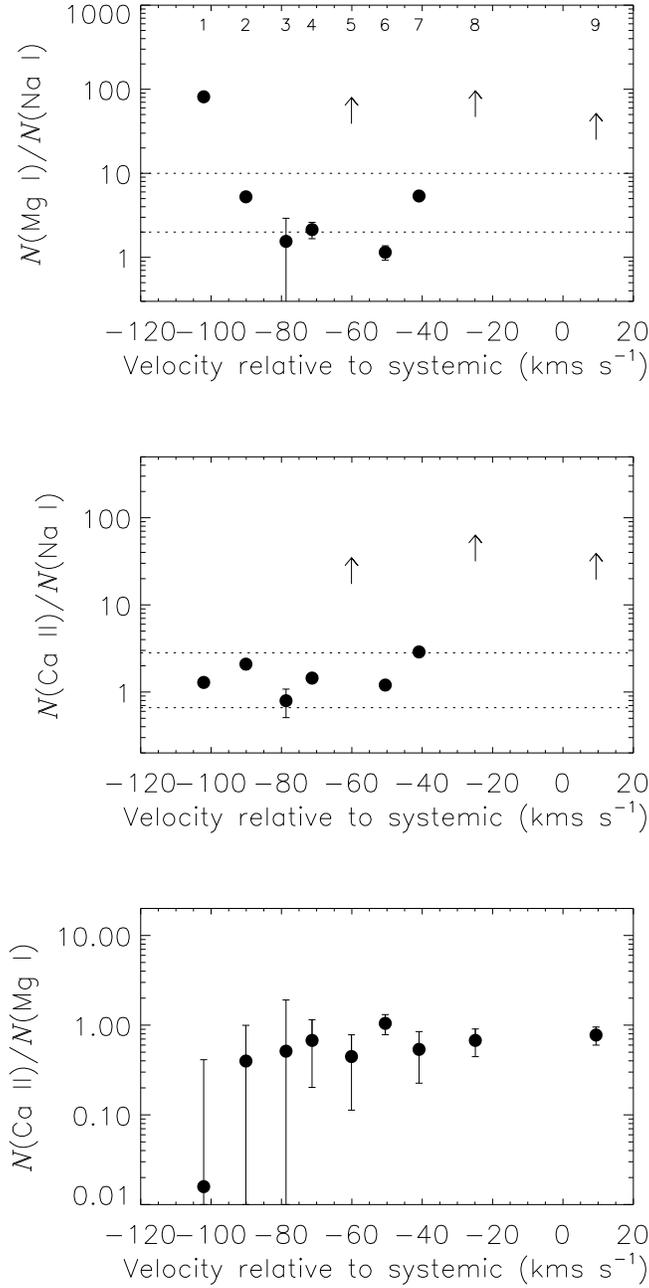,height=18cm,angle=0}}
\caption{Plots of the derived column density ratios for the 9 low-ionization
components seen in NGC~3877. The 
dotted lines in the top figure represent the range of values of
$N$(Mg~I)/$N$(Na~I) seen by \cite{pett77} towards Galactic stars.
The dashed lines in the middle figure represent the range of 
$N$(Ca~II)/$N$(Na~I) seen in the Milky Way
for absorbing components at peculiar velocities of $\Delta v\:\leq\:50$~\kms\ 
(\cite{semb94}).  The corresponding component number given in Table~3 is shown top
of upper panel. \label{ratio.fig}}
\end{figure}

In Figure~\ref{ratio.fig} we show the
$N$(\ion{Mg}{1})/$N$(\ion{Na}{1}), $N$(\ion{Ca}{2})/$N$(\ion{Na}{1}),
and $N$(\ion{Ca}{2})/$N$(\ion{Mg}{1}) ratios from the fits in
Table~3. The use of the first of these,
$N$(\ion{Mg}{1})/$N$(\ion{Na}{1}), as a diagnostic of gas temperature
has been discussed by \cite{pett77}, who showed that low ratios in the
Galactic disk could be explained if absorption occurred only in cool
($T\:\leq\:80$~K) \ion{H}{1} clouds. The ratios found towards SN~1998S
show a mixture of values, suggesting both cool quiescent clouds and
warmer halo structures. Similarly, the
$N$(\ion{Ca}{2})/$N$(\ion{Na}{1}) ratio, usually interpreted as an
indicator of the calcium gas-phase abundance, also shows a wide range;
both, however, suggest a difference in physical conditions above and
below $\sim\:-50$~\kms . We consider these ratios more fully below.


\subsubsection{Interpretation}

At the position of the SN sightline, the observed H~I from 21~cm
emission in NGC~3877 is measured from $730-910$~\kms\ (\cite{broe94}),
or $\Delta v\:=\:v - v_{\rm{systemic}}$ of $-180$ to 0~\kms.  The
galaxy is highly inclined, $i\:=\:84\arcdeg$, yet the distance of the
SN from the major axis is still non-negligible, which suggests that
the true radial distance, $R$, of the object from the center of the
galaxy is large, as much as 10.3~kpc, assuming it lies in the galactic
plane. If so, the SN is near the edge of the H~I radius of the galaxy,
measured to be $\sim 10$~kpc (\cite{sand98}). The question remains,
however, whether the SN sits at the front or back of the disk, and
whether the low-ionization absorption can be understood in terms of
disk-gas along the line of sight. If the true distance of the SN in
the disk is 10~kpc, then the de-projected angle between it and the
major axis of the galaxy is $\simeq\:71\arcdeg$. The true radial
velocity, $v(R)$, of gas in the disk of NGC~3877 at 10~kpc is 175~\kms
, as measured from the rotation curve (\cite{sand98}).  Hence the
projected velocity, $v_o$, along the line of sight of gas close to the
supernova's environment would be $v_o - v_{\rm{sys}}\:=\:v(R)\sin
i\cos\theta\:\simeq\:-56$~\kms , where $\theta$ is the angle between a
point in the disk along the sightline and the center of the disk. This
is true whether the SN is in the edge of the disk closest or furthest
from us, and the velocity lines up well with the red edge of the
\ion{Na}{1} complex in Figure~\ref{vel1b.fig}. Perhaps the simplest
argument to suggest that the SN must lie on the trailing side of the
galaxy is to point out that at 10~kpc, the rotation curve is flat, and
were the SN to lie on the leading edge, {\it all} the disk gas which
lies between the probe and us would essentially be at the same
velocity, $\simeq\:-56$~\kms . From Figure~\ref{vel1b.fig}, this does
not appear to be the case. If the galaxy were exactly edge-on to us,
and the SN occurred at the back of the disk, we might expect to see
absorption from gas through the entire disk. The most negative
velocity gas would occur where the sightline passes closest to the
center of the galaxy (i.e. $\theta\:=\:0$). This value (essentially
the projected distance of the SN from galaxy center) is $3.5$~kpc ,
and $v(R)$ is still $\sim\:150$~\kms. The projected velocity of the
gas would then be $-146$~\kms\ from systemic, and we do not see any
absorption at these velocities. In fact, given that the galaxy is {\it
not} edge-on, and assuming that the H~I disk has a height similar to
that in our own Galaxy, $\sim\:1$~kpc, the sightline never intercepts
gas near the center of the galaxy --- it probes the ISM over perhaps
only half a disk radius, depending on where exactly the supernova sits
at the back of the disk. Then for $R=5$~kpc, and $v(R)\:=\:160$~\kms,
$v_o - v_{\rm{sys}}\:=\:-90$~\kms , which is, remarkably, the same as
the most blueward non-variable component we see.

If this simple model of absorption by the back quarter or so of the
galactic disk is correct, then it also explains to a large
degree the distribution of column density ratios vs.\ velocity as
plotted in Figure~\ref{ratio.fig}. The components between $-40$ and
$-90$~\kms\ (i.e. numbers $2-4$ and $6-7$) all lie in the disk of the
galaxy and have $N$(\ion{Ca}{2})/$N$(\ion{Na}{1}) and
$N$(\ion{Mg}{1})/$N$(\ion{Na}{1}) ratios similar to those seen in the
disk of the Milky Way. Two other components, numbers 8 and 9, have
velocities which do not fit in with the rotation of the galaxy ---
these components have high {\it peculiar} velocities, relative to the
gas in the disk, similar to those seen in our own galaxy, and have
correspondingly high $N$(\ion{Ca}{2})/$N$(\ion{Na}{1}) and
$N$(\ion{Mg}{1})/$N$(\ion{Na}{1}) values. Component 5 would then be
interpreted perhaps as arising from a warmer intercloud medium.  This
just leaves us to explain the origin of the variable component~1,
which has no apparent relation to the co-rotating ISM gas in
NGC~3877. This component corresponds to many other unusual features,
however, which we now discuss.


\subsection{Circumstellar absorption and emission from SN~1998S}

\begin{figure}
\centerline{\psfig{figure=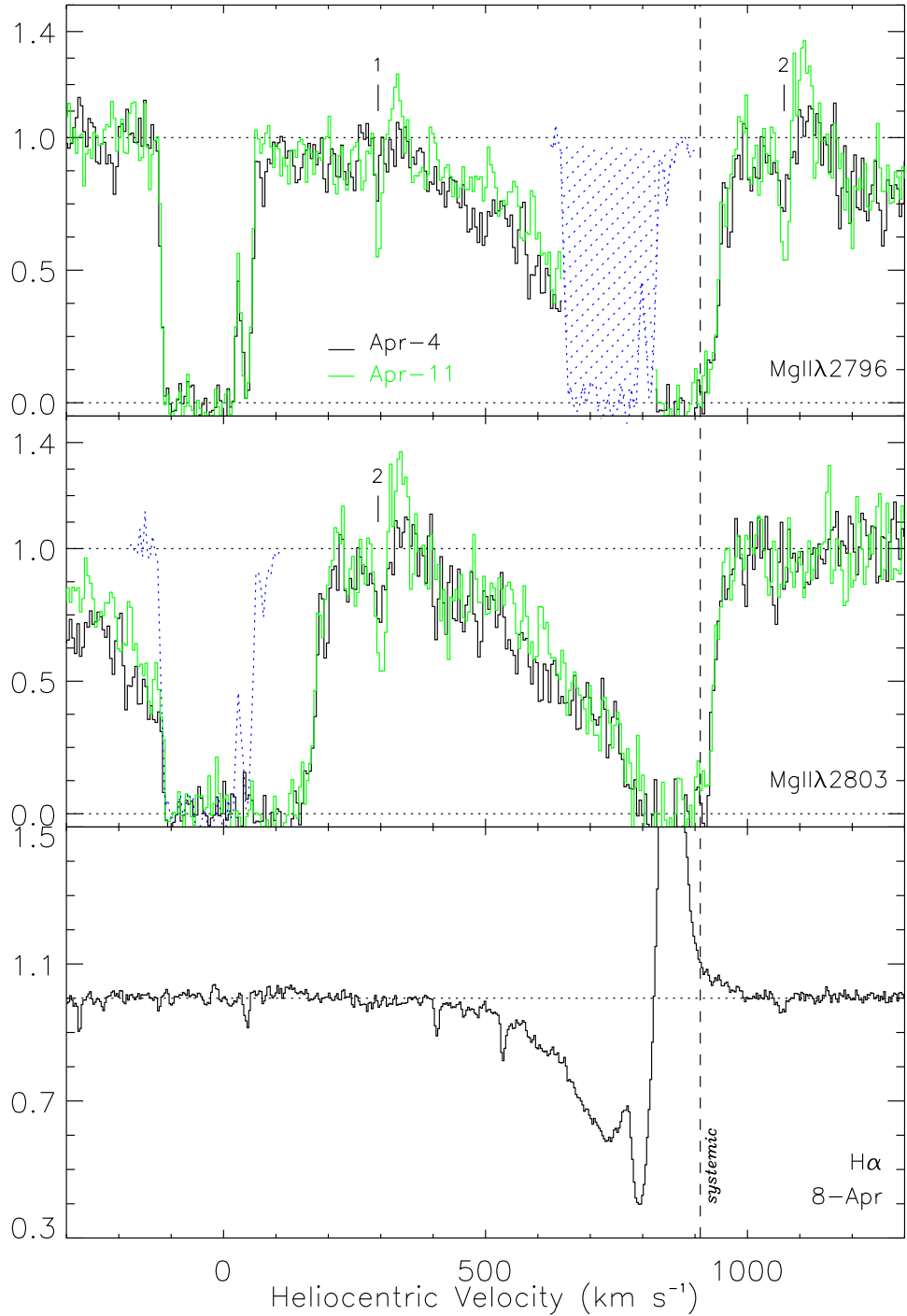,height=20cm,angle=0}}
\figcaption{Upper two panels show the Mg~II line profiles
obtained with STIS on April-4 and April-11.  
See text for an explanation of the hatched
areas. Bottom panel
shows the optical spectra covering the emission and absorption from
H$\alpha$.  Mg~II $\lambda 2791.60$ (labelled `1')
and Mg~II $\lambda\lambda2798.754,2798.823$ (`2') are also seen, 
arising at the same
velocity as the variable, narrow Na~I component No.~1 shown in
Figure~\ref{exg.fig}. \label{vel2.fig} }
\end{figure}

Although the initial aim of our observations was to examine the UV
absorption from the interstellar medium of NGC~3877, our data have
revealed unusual absorption line profiles which have more to do with
the environment of the supernova itself. In Figure~\ref{vel2.fig} we
show the Mg~II absorption both from our own Galaxy and from
NGC~3877. The system is complicated by Mg~II $\lambda 2796$ at the
velocity of NGC~3877 being contaminated with Mg~II $\lambda 2803$ from
our Galaxy at rest. To highlight the possible degree of contamination,
we have shifted and overplotted the Galactic Mg~II $\lambda 2796$
profile at the position expected for Galactic Mg~II $\lambda 2803$, as
shown by the hatched region in the top panel. Hence the absorption
seen in this region is a superposition of saturated lines from both
Mg~II $\lambda 2796$ absorption in NGC~3877 and Mg~II $\lambda 2803$
from the Milky Way. This same blend of absorption is also seen in the
middle panel, with Mg~II $\lambda 2803$ from the Galaxy now at rest,
but contaminated with Mg~II $\lambda 2796$ from NGC~3877.


Despite this confusion, it is clear that there exists a `shelf' of
absorption spanning $\sim\:350$~\kms\ in addition to the saturated
components expected from the ISM of NGC~3877 itself at
$v_\odot\:=\:800-900$~\kms . We see a smooth decline in the strength
of the absorption as the velocity from systemic increases.  There is
also some evidence that the strength of the absorption in this region
has declined between April-4 and April-11, although we note that
defining the continuum in regions of wide absorption is difficult.

\begin{figure}
\centerline{
\psfig{figure=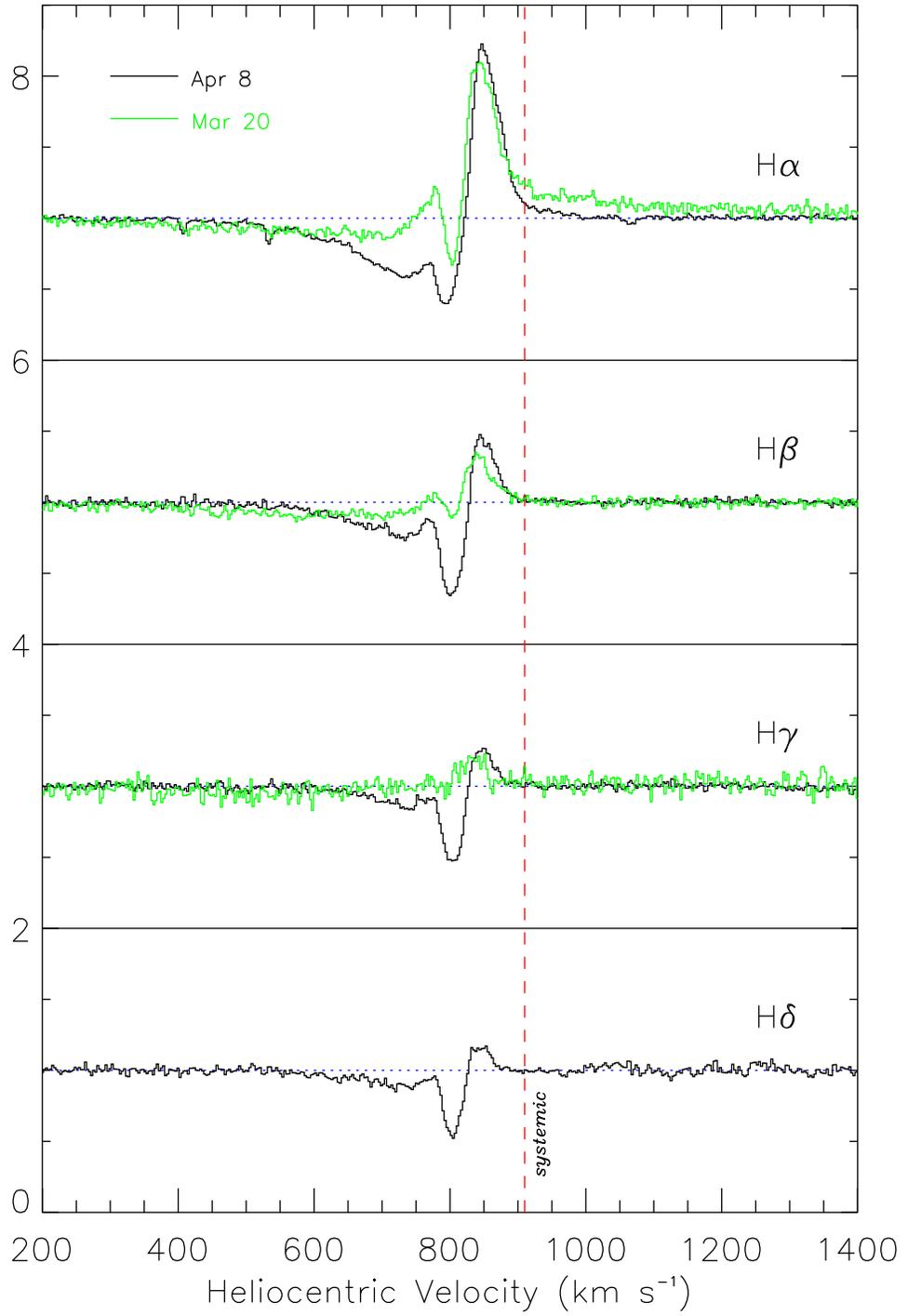,height=20cm,angle=0}}
\figcaption{\label{timecf_balmer.fig}
Sequence of Balmer lines observed 19 days
apart with the UES at the \WHT. }
\end{figure}

Features in the optical data also show unusual conditions. At the
bottom of Figure~\ref{vel2.fig} we overplot the H$\alpha$ profile
taken April-8th. The `shelf' seen in Mg~II $\lambda\lambda 2796,2803$
can also be seen in H$\alpha$ absorption, but now part of a classic
P-Cygni profile defined by H$\alpha$ emission to the blue. This
absorption can be seen in other Balmer lines too, in H$\alpha$,
H$\beta$, H$\gamma$ and H$\delta$. The absorption lines do not result
from incorrect background subtraction of emission from other parts of
the galaxy filling the slit---extraction of the supernova signal
without background subtraction also shows the P-Cygni profiles.  We
show these profiles in Figure~\ref{timecf_balmer.fig}, with the data
from the two different epochs plotted over each other. (H$\epsilon$ is
also present, but blended with Ca~II~H from the ISM of NGC~3877, so is
not shown in Figure~\ref{timecf_balmer.fig}.) There
is no doubt that the absorption complex has {\it strengthened}
significantly between March-20 and April-8, in contrast to the
\ion{Mg}{2} which has declined. This suggests that the gas forming the
\ion{Mg}{2} has increased in ionization, with more of the magnesium
forming \ion{Mg}{3}. 


The exact level of H$\alpha$ taken
March-20 is uncertain, because no flux calibrators are available, and
there is difficulty in defining where best to fit the continuum. There
is, however, a possibility that the H$\alpha$ shows weak excess emission
as a wide red wing covering $\sim 600$~\kms\ on March-20, emission
which has disappeared by April-8. These P-Cygni profiles, so well
displayed in Balmer lines, can also be seen in many He~I lines
(e.g. He~I $\lambda5875$ in Figure~\ref{vel1b.fig}); we list all the
identified emission lines in April-8's data in
Table~4, and note whether the lines show the P-Cygni
profile. Some do not, and these emission lines are likely to be due to
either the H~II region in which the supernova exploded (see below), or
from other emission line regions along the line of sight not directly
connected with the supernova's environment.


\begin{table}[t]
\begin{small}
\begin{center}
\tablewidth{6cm}
\begin{tabular}{lc}
\multicolumn{2}{c}{Table 4. 
Emission lines identified}\\
\multicolumn{2}{c}{in 8-Apr-1998 optical data}\\
\hline\hline
Line & P-Cygni \\
\hline
\ion{[Ne}{3}] $\lambda 3867$	& n   \\
\ion{He}{1} $\lambda3870$ 	& n   \\
\ion{He}{1} $\lambda3887$ + H 8 & y   \\
\ion{He}{1} $\lambda3963$	& y   \\
H$\epsilon$			& y   \\
\ion{He}{1} $\lambda 4025$	& y   \\
H$\delta$, H$\gamma$            & y   \\
\ion{[O}{3}] $\lambda4361$	& n   \\
\ion{He}{1} $\lambda4470$	& y   \\
\ion{He}{1} $\lambda4711$	& y   \\
\ion{He}{1} $\lambda4290$	& ?   \\
\ion{[Fe}{3}] $\lambda4656$	& n   \\
H$\beta$			& y   \\
\ion{[O}{3}] $\lambda4957$	& n   \\
\ion{[O}{3}] $\lambda5005$	& n   \\
\ion{He}{1} $\lambda5014$	& y   \\
\ion{[Fe}{3}] $\lambda5268$	& n   \\
\ion{[N}{2}] $\lambda5753$	& n   \\
\ion{He}{1} $\lambda5889$	& y   \\
H$\alpha$			& y   \\
\ion{[N}{2}] $\lambda6581$:     & n   \\
\ion{O}{1} $\lambda7769$		& n   \\
\hline
\end{tabular}
\end{center}
\end{small}
\end{table}

What is also clear from these P-Cygni profiles is that there exists
strong, narrow absorption in addition to the weaker, wider shelf
already described, located at the bottom of the P-Cygni
trough. Although the exact center of this absorption is hard to
measure, because it is superimposed on the emission line, it seems
clear that it lies at the same velocity as the variable Na~I component
at $v_o - v_{\rm{sys}}\:=\:-102$~\kms\ discussed in \S2.2. This can
best be seen in Figure~\ref{vel1b.fig} where the sharp absorption in
He~I $\lambda5875$ is coincident with the varying Na~I
component. Further, there are other absorption lines in our \HST\
spectra which coincide with this varying component. For example, there
are the rarely seen Mg~II $\lambda 2791.60$ and
Mg~II $\lambda\lambda2798.754,2798.823$ lines (labelled `1' and `2' in
Figure~\ref{vel2.fig}) which arise at this same velocity.

\begin{figure}
\vspace*{-1cm}
\hspace*{1.6cm}\psfig{figure=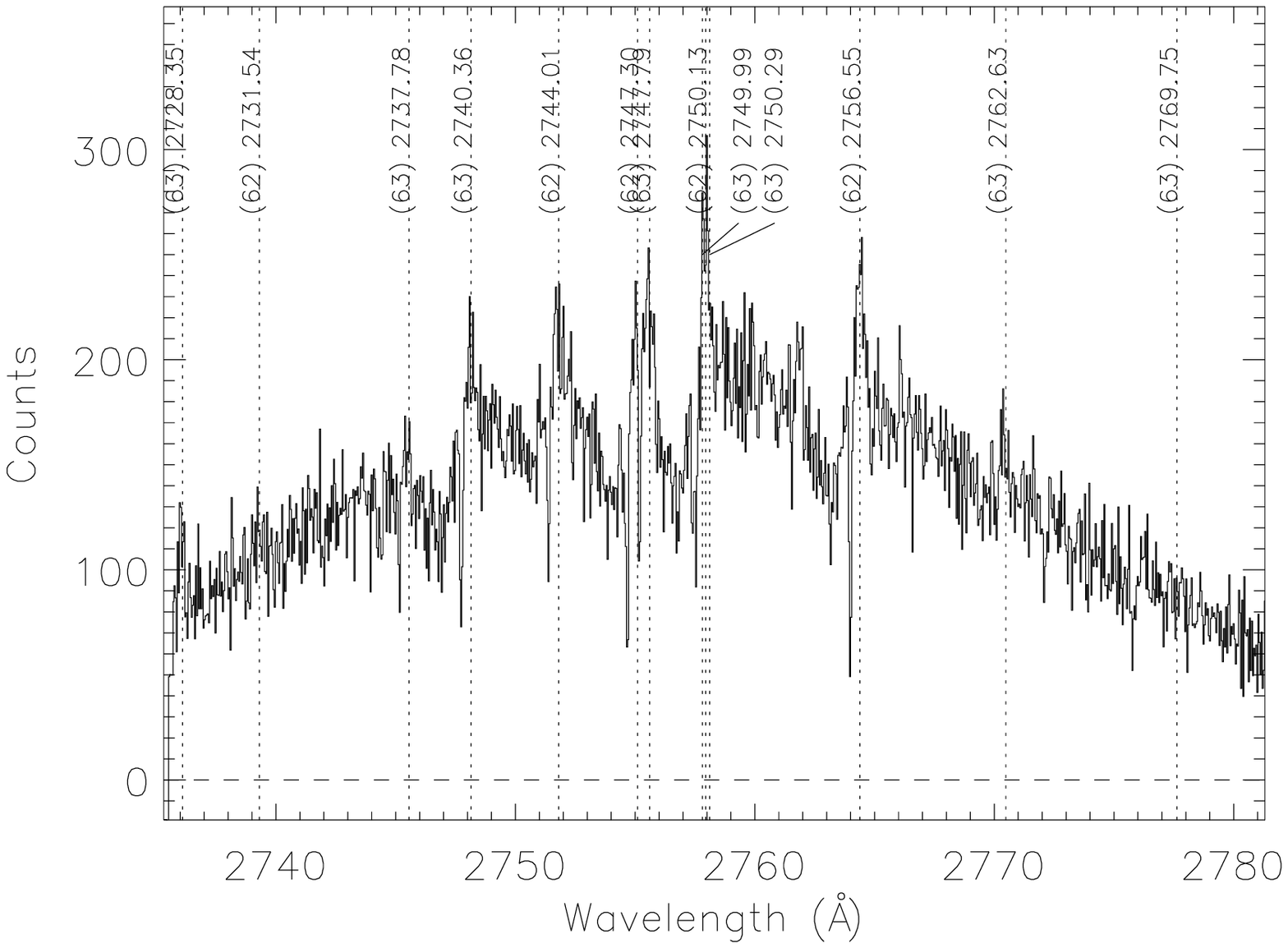,height=10cm,angle=0}
\vspace*{-0.5cm}
\hspace*{1.6cm}\psfig{figure=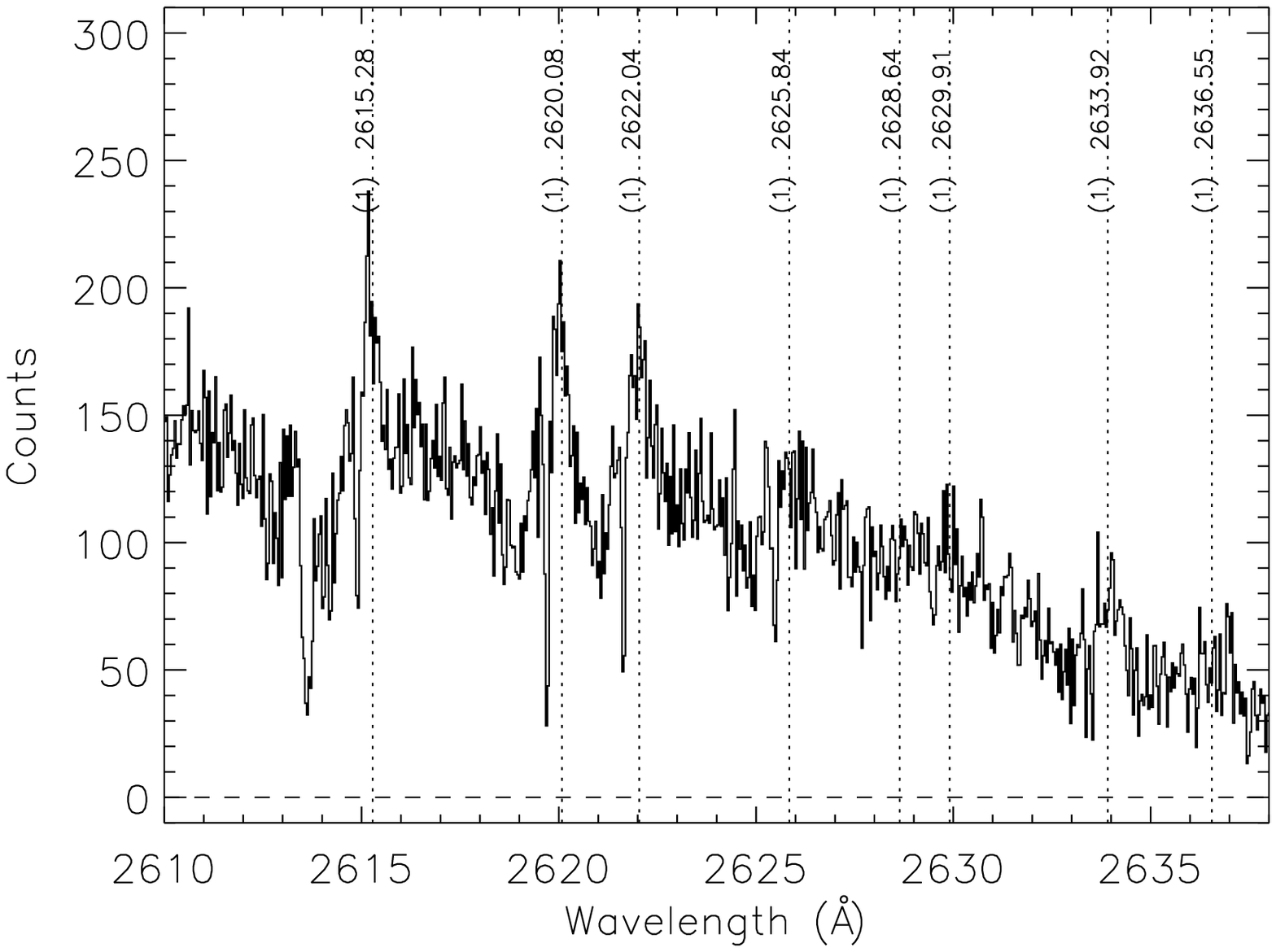,height=10cm,angle=0}
\figcaption{\label{fe2lines.fig}
Two figures showing some of the metastable Fe~II lines detected
in April-4's STIS spectra. All the lines arising from the uv62, uv63
(top) and uv1 (bottom) transitions are plotted over the wavelength
ranges shown, shifted by 850~\kms , regardless of whether we consider
them detected. Also shown are the multiplet number and the rest
wavelength of the line. The uv62 and uv63 lines involve a change from a
$a^4D$ to a $z^4F^o$ state, while the uv1 lines arise from a change in
energy from $a^6D$ to $z^6D^o$.}
\end{figure}

Even more unusual are the many uv1, uv2, uv62, uv63, uv158 and uv159
multiplet Fe~II lines seen in the STIS spectra, also with P-Cygni
profiles.  All the transitions we detect arise from the $a^4H$,
$a^6D$, and $a^4D$ states.  Examples of these lines are shown in
Figure~\ref{fe2lines.fig}, where we have plotted all the lines of a
given multiplet shifted from rest to a velocity of 850~\kms .  Both
the broad shelf and the narrow absorption is seen in these profiles,
as detected in the other species. There is no clear difference between
the profiles between April-4 and April-11. Although the data are of
too low a signal-to-noise, and the continuum of the P-Cygni profiles
too ill-defined, to measure physical parameters accurately, the narrow
component is clearly resolved in the data. Crudely, the lines appear
to be a factor of 2 wider than the instrumental resolution, which
gives a lower limit of $b\:>\:6$~\kms, assuming that the spectra have
a resolution of 1.7 pixels FWHM, as given by the STIS Instrument
Handbook. This is much wider than that deduced from the Na~I component
alone, ($b\:\simeq\:2.7\pm0.2$~\kms, Table 3) which suggests a much
more complicated structure to the narrow component than a single
absorbing cloud.


It seems clear that all these features --- the emission lines and the
narrow, variable absorption component superimposed on a broader shelf
of absorption --- are unlikely to have anything to do with the
supernova ejecta itself. Typically, ejecta material is moving at
$\sim10^{4}$~\kms\ at the epochs at which we observed SN~1998S, hence
the P-Cygni profiles would be considerably wider than the $\sim
30$~\kms\ FWHM observed here. Indeed, the far-UV G140L spectra show
the expected broad P-Cygni profiles from the expanding photosphere of
the supernova in a variety of ions.

Instead, it seems likely that the
P-Cygni profiles arise from circumstellar material, blown off from the
progenitor star.
Such an interpretation is not unprecedented.  Narrow Balmer lines and
\ion{He}{1} $\lambda5876$ have been seen before in a supernova
spectrum, towards SN~1984E in NGC~3169 (\cite{dopi84}). Their
detection was cited as evidence of a superwind flowing out from the
precursor star. In this case, the wind had a velocity of some
3000~\kms , much greater than we observe towards SN~1998S.  UV
emission lines towards SN~1979C in M~100 were also interpreted as
being due to a UV-emitting shell of compressed gas from pre-existing
circumstellar material (\cite{pana80}), and indeed, a narrow H$\alpha$
P-Cygni profile may also have been seen (\cite{bran81}). Again,
however, the width of these UV emission lines ranged from being
unresolved at a limit of 1000~\kms\ to a maximum of 4000~\kms, larger
than we see towards SN~1998S.

The circumstellar absorption detected towards SN~1998S is obviously
far more quiescent than these superwinds.  Stellar winds from, e.g.,
O- \& B-type stars, or Population~I Wolf-Rayet stars, are known to
have high mass-loss rates and strong outflows, giving rise to the
classic P-Cygni profiles. However, these outflows move with a speed of
a few $\times10^{3}$~\kms\ (see, e.g. \cite{lozi92}), again, much
higher than we observe towards SN~1998S. Perhaps the most interesting
comparison is the detection of the same Fe~II UV multiplets seen in
the winds of A-type supergiants (\cite{tala87};
\cite{verd99}). Although not all A-type stars show these absorption
lines, those that do are characterized by outflows moving at
only a few hundred ~\kms, along with variability in the absorption. It
seems likely that this type of outflow is responsible for the
complicated emission and absorption seen in NGC~3877.

Another possibility is that at least some component of the absorption
arises from a proto-planetary disk surrounding the
progenitor. Observations towards $\beta$-Pictoris and other A-type and
late O-type stars show the same \ion{Fe}{2} metastable lines as we find
towards SN~1998S, as well as variable absorption (e.g. \cite{lagr96},
\cite{kond85}, \cite{chen97}). If so, the absorption we see might well
mark the presence of an early solar system shortly before its
destruction by the supernova. The most obvious argument against this
interpretation is simply that the absorbing shell or wind is probably
much further away from the progenitor than would be expected from a
proto-planetary disk. The existence of the \ion{Fe}{2} lines in our
last STIS spectra taken April-11 means that the ejecta from the
supernova had not yet reached the absorbing material, some 39 days
after the supernova was detected. Assuming simply that the ejecta
moved with a constant velocity after outbreak, at, e.g., 5,000~\kms,
then for the absorbing material to remain undisturbed (i.e, assuming
that the ejecta would have catastrophic effects on the absorbing
material), the distance of the material from the progenitor would be
$>\: 100$~AU. This is much larger than the few AU postulated for the
$\beta$-Pictoris disk. 

We also note that the shell has a velocity relative to the progenitor
in our simple model of where the supernova sits in the galaxy,
discussed in \S2.2.2. We associated the progenitor to be at a radius
in the galaxy where the ISM gas has a projected velocity along the
line of sight to be $v-v_{\rm{systemic}}\:\approx\:-50$~\kms ; if this
gas forms a reference frame for the progenitor, then the variable
absorption occurring at $v-v_{\rm{systemic}}\:\approx\:-100$~\kms\ is
$\approx\:50$~\kms\ different, a value which could therefore be taken
to be the shell's speed. This, combined with the broader absorption
shelf, would suggest a dense shell with a speed of about
50~\kms\ and a more highly ionized shell
moving at $\sim\:300$~\kms .

There are obvious complications in trying to interpret the observed
data with stellar outflows. First, the material does not necessarily
form a single bubble of gas; towards SN~1987A, for example, it is
believed that the slowly expanding wind from the red supergiant phase
of the progenitor is being caught by a faster wind from its blue
supergiant phase, producing two shock fronts. There is also the
obvious fact that the wind must hit the ambient ISM surrounding the
progenitor, producing another type of shock front. These process are
not well understood in our own Galaxy, and applying them to the
features observed towards SN~1998S is clearly not easy. Second,
towards a supernova, the ionization structure of the gas is
complicated by the addition of the UV flash occurring at the initial
explosion of the supernova. The effects of this ionizing source are
superimposed on the ionization of the winds and shells caused by the
progenitor and the earlier mechanical interactions. Third, in
attempting to link the variability of the narrow absorption with the
types of variations seen in the outflows of type-A supergiants, there
is the added complication that the UV source against which the
absorption is seen is expanding. This could in principle
produce---apparently---variable absorption features, depending on the
density structure within the outflow (whether clumps exist in the
outflow) and the covering factors of the clumps. Finally, although
the Fe~II lines can be found in the spectra of A-type stars and their
circumstellar material, they rarely show the well defined P-Cygni
profiles we see towards SN~1998S.

Hence, although we believe that we are seeing unique signatures of the
outflows from the supergiant progenitor of SN~1998S, the exact details
of the structure of the circumstellar material are beyond the scope of
this paper.

\section{Summary}

We have observed SN~1998S which exploded in NGC~3877 with the UES
(6$-$7~\kms\ FWHM) at the \WHT\ and with the E230M echelle of the STIS
(8~\kms\ FWHM) aboard \HST.  Both data sets were obtained at two
epochs, separated by 19 (optical) and seven days (\HST\ data) (see
Table~1). We summarize our results as follows. From our own Galaxy we
detect:

\begin{itemize}

\item interstellar \ion{Ca}{2}~K, \ion{Fe}{2} $\lambda\lambda
2600,2586,2374,2344$, \ion{Mg}{1} $\lambda2852$, and probably
\ion{Mn}{2} $\lambda 2576$, absorption lines at
$v_{\rm{LSR}}\:=\:-95$~\kms\ arising from the outer edge of the High
Velocity Cloud Complex~M. We derive gas-phase abundances of
[Fe/H]$\:=\:-1.4$ and [Mn/H]$\:=\:-1.0$, values which are very similar
to warm disk clouds found in the local ISM.

\end{itemize}

This is the first detection of manganese from a Galactic
HVC, and we believe that the values of the iron and manganese 
gas-phase abundances argue
against the HVC material having an extragalactic origin.

At the velocity of NGC~3877 we detect:

\begin{itemize}

\item interstellar \ion{Mg}{1} $\lambda 2852$,
\ion{Mn}{2} $\lambda\lambda 2576, 2594,2606$, \ion{Ca}{2}~K and
\ion{Na}{1}~D$_2$,D$_1$ absorption lines, spanning a velocity range of
$-102$ to $+9$~\kms\ from the systemic velocity of the galaxy
(910~\kms ). In particular, the component at $-102$~\kms\ is seen to
increase by a factor of $\apg\:1$ dex in $N$(\ion{Na}{1}) between
March-20 and April-8, and in $N$(\ion{Mg}{1}) between April-4 and
April-11;

\item Balmer and \ion{He}{1} P-Cygni profiles in the optical,
with a narrower absorption component superimposed at the bottom of the
absorption trough of the profile. Both the broad and narrow components
are seen to vary substantially in their absorption strength between
March-20 and April-8; 

\item a broad shelf of \ion{Mg}{2} $\lambda 2796,2803$ absorption
spanning a velocity range of $~\sim 350$~\kms, with an apparent
optical depth gently declining towards more negative velocities. This
shelf covers the same velocity extent as the Balmer and He~I lines in
the optical (which form the blue side of a P-Cygni profiles) but there
is no evidence of any narrow emission forming a \ion{Mg}{2}
P-Cygni profile. There is some suggestion that this shelf has {\it
decreased} in strength over seven days between April-4 and April-11;

\item metastable \ion{Fe}{2} absorption lines from the UV multiplets
uv1, uv2, uv62, uv63, uv158 and uv159. These too show the same P-Cygni
profiles seen in the optical, and are again superimposed with the narrow
component at the trough of the broad absorption part of the profile.
These lines do not obviously vary over the seven days between the two
STIS observations;

\item rarely seen \ion{Mg}{2} $\lambda 2791$ and
\ion{Mg}{2} $\lambda\lambda 2798.75,2798.82$ lines. These show P-Cygni
profiles on April-11 which do not exist on April-4.

\end{itemize}

Most of the low-ionization absorption (the \ion{Ca}{2}, \ion{Na}{1},
\ion{Mg}{1}, \ion{Mn}{2} lines) can be understood in terms of gas
co-rotating with the disk of NGC~3877, providing the supernova is at
the back of the disk as we observe it, and the H~I disk is a few
kpc thick. However, the variable component seen in all the other
lines, and the accompanying emission which goes to form the classic
P-Cygni profiles, most likely arises in slow moving circumstellar
outflows originating from the red supergiant progenitor of SN~1998S.

\acknowledgments

We would like to thank Steve Smartt for taking the UES Service data on
April-8, and in particular Sandi Catalan and Janet Wood for using
their UES observing time on March-20. Thanks also to Peter Garnavich
for sharing initial low resolution observations, Steve Bennett \&
Alfonso Aragon-Salamanca for measuring SN magnitudes shortly after
detection, Stefi Baum for sheparding the Phase 2 program at STScI, and
the referee, Steven Federman, for a careful reading of our
paper. The Fe~II P-Cygni profiles listed in Table~4 were initially
identified using Peter van Hoof's {\it Atomic Line List v2.02} web
page (www.pa.uky.edu/\~{}peter/atomic/index.html).

Support for this work was provided by NASA through grant numbers
GO-06728.01-95A and GO-06707.01-95A from the Space Telescope Science
Institute, which is operated by the Association of Universities for
Research in Astronomy, Inc., under NASA contract NAS5-26555.


\begin{thebibliography}{}


\bibitem[Anders \& Grevesse 1989]{and89}Anders, E., \& Grevesse, N. 1989, 
Geochim.\ Cosmochim.\ Acta, 53, 197



\bibitem[Blades et al.\ 1988]{blad88} Blades, J. C., Wheatley, 
J. M., Panagia, N., Grewing, M., Pettini, M. \& Wamsteker, W. 1988, \apj, 
334, 308 

\bibitem[Bowen 1991]{bowen91}Bowen, D. V. 1991, MNRAS, 251, 649

\bibitem[Bowen et al.\ 1994]{bowen94} Bowen, D. 
V., Roth, K. C., Blades, J. C.  \& Meyer, D. M. 1994, \apjl, 420, L71 

\bibitem[Bowen et al.\ 1995]{bowen95}Bowen, D. V., Blades, J. C., \& Pettini,
M. 1985, \apj, 448, 634


\bibitem[Blitz et al.\ 1999]{bli99}Blitz, L., Spergel, D. N., Teuben,
P. J., Hartmann, D., \& Burton, W. B. 1999, \apj, 514, 818

\bibitem[Branch et al.\ 1981]{bran81} Branch, D., Falk, S. W., 
Uomoto, A. K., Wills, B. J., McCall, M. L. \& Rybski, P. 1981, \apj, 244, 
780 

\bibitem[Broeils \& van Woerden 1994]{broe94}
Broeils, A. H. Van Woerden, H. 1994 A\&AS, 107, 129

\bibitem[Cheng et al.\ 1997]{chen97} Cheng, K.-P., Bruhweiler, F. C.,
\& Neff, J. E. 1998, \apj, 481, 866

\bibitem[Danly et al.\ 1993]{danl93}
Danly, L., Albert, C. E., \& Kuntz, K. D. 1993, \apj, 416, L29

\bibitem[D'Odorico et al.\ 1985]{dodo85}D'Odorico, S., 
Pettini, M. \& Ponz, D. 1985, \apj, 299, 852 

\bibitem[Dopita et al.\ 1984]{dopi84} Dopita. M. A., Evans, R., Cohen,
M., \& Schwartz, R. D. 1984, \apj, 287, L69

\bibitem[Hartmann \& Burton 1997]{hart97}
Hartmann, D. \& Burton, W. B. 1997, Atlas of galactic neutral hydrogen,
Cambridge University Press.

\bibitem[Jenkins et al.\ 1984]{jenk84}Jenkins, E. B., Rodgers, 
A. W., Harding, P., Morton, D. C. \& York, D. G. 1984, \apj, 281, 585 

\bibitem[Keenan et al.\ 1995]{keen95}
Keenan, F. P., Shaw, C. R., Bates, M., Dufton, P. L., \& Kemp, S. N. 1995,
\mnras, 272, 599

\bibitem[Lagrange et al.\ 1996]{lagr96} Lagrange et al.\ 1996, \aa,
310, 547

\bibitem[Lozinskaya 1992]{lozi92} Lozinskaya, T. A. 1992, in Supernovae
and Stellar Wind in the Interstellar Medium, (American Institute of
Physics), p 279.

\bibitem[Lu, Sargent, \& Barlow (1996)]{lu96}Lu, L., Sargent, W. L. W., \&
Barlow, T. A. 1996, \apjs, 107, 475

\bibitem[Lu et al.\ (1995)]{lu95}Lu, L., Savage, B. D., Tripp, T. M., 
\& Meyer, D. M. 1995, \apj, 447, 597

\bibitem[Meyer et al.\ (1995)]{mey95}Meyer, D.M., Lanzetta, K.M., \& Wolfe, A.M. 1995, \apj, 451, L13

\bibitem[Meyer \& Roth 1991]{meye91}Meyer, D.M., \& Roth,
K. C. 1991, \apj, 383, L41

\bibitem[Panagia et al.\ 1980]{pana80} Panagia, N. et al.\ 1980, \mnras,
192, 861

\bibitem[Pettini et al.\ (1977)]{pett77} Pettini, M., Boksenberg, 
A., Bates, B., McCaughan, R. F. \& McKeith, C. D. 1977, \aap, 61, 839 

\bibitem[Pettini et al.\ 1982]{pett82} Pettini, M., et al. 
1982, \mnras, 199, 409 

\bibitem[Pettini et al.\ (1999)]{pet99}Pettini, M., Ellison, S. L.,
Steidel, C. C., \& Bowen, D. V. 1999, \apj, 510, 576

\bibitem[Roth \& Songaila (1999)]{rot99}Roth, K. C. \& Songaila,
A. 1999, in preparation

\bibitem[Ryans et al.\ 1997]{ryan97}
Ryans, R. S. I., Keenan, F. P., Sembach, K. R., \& Davies, R. D. 1997, \mnras,
289, 83


\bibitem[Ryan, Norris, \& Beers (1996)]{rya96}Ryan, S. G., Norris, J. E.,
\& Beers, T. C. 1996, \apj, 471, 254

\bibitem[Savage \& Sembach (1996)]{sav96}Savage. B. D. \& Sembach,
K. M. 1996, \araa, 34, 279

\bibitem[Sanders \& Verheijen 1998]{sand98}
Sanders, R. H. \& Verheijen M. A. W. 1998, ApJ, 503, 97

\bibitem[Sembach~et~al.~1993]{semb93} Sembach, K. R.,
Danks, A. C., \& Savage, B. D. 1993, A\&AS, 100, 107

\bibitem[Sembach \& Danks 1994]{semb94} Sembach, K. R. \& 
Danks, A. C. 1994, \aap, 289, 539 

\bibitem[Steidel et al.\ 1990]{stei90}Steidel, C. 
C., Rich, R. M.  \& McCarthy, J. K. 1990, \aj, 99, 1476 

\bibitem[Steidel et al.\ (1995)]{ste95}Steidel, C. C., Bowen, D. V.,
Blades, J. C., \& Dickinson, M. 1995, \apj, 440, L45

\bibitem[Talavera \& G\'{o}mez de Castro 1987]{tala87} Talavera, A., 
\& G\'{o}mez de Castro, A. I. 1987, \aa, 181, 300

\bibitem[Verdugo et al.\ 1999]{verd99} Verdugo, E., Talavera, A., \&
G\'{o}mez de Castro, A. I., 1999, A\&AS, 137, 351

\bibitem[Vidal-Madjar et al.\ 1987]{vida87}Vidal-Madjar, A., 
Andreani, P., Cristiani, S., Ferlet, R., Lanz, T. \& Vladilo, G. 1987, 
\aap, 177, L17 

\bibitem[Vladilo 1998]{vla98}Vladilo, G. 1998, \apj, 493, 583

\bibitem[Kondo \& Bruhweiler 1985]{kond85} Kondo, Y., \& Bruhweiler,
F. C. 1985, \apj, 291, L1

\bibitem[Wallerstein et al.\ 1972]{wall72} Wallerstein, G., Conti,
P. S., Greenstein, J. L. 1972, Astrophysical Letters, 12, 101

\bibitem[Wakker \& van Woerden 1991]{wakk91}
Wakker, B. P.  \& van Woerden, H. 1991, A\&A, 250, 509

\bibitem[Wakker \& van Woerden 1997]{wakk97}
Wakker, B. P.  \& van Woerden, H. 1997, \araa, 35, 217


\end{thebibliography}
\end{document}